\documentclass[fleqn,usenatbib]{mnras}

\usepackage{newtxtext,newtxmath}

\usepackage[T1]{fontenc}

\DeclareRobustCommand{\VAN}[3]{#2}
\let\VANthebibliography\thebibliography
\def\thebibliography{\DeclareRobustCommand{\VAN}[3]{##3}\VANthebibliography}

\usepackage{graphicx}	
\usepackage{amsmath}	

\newcommand{\NaD}{\mbox{Na \textsc{i d}}}
\newcommand{\MgI}{\mbox{Mg b}}
\newcommand{\OIII}{\mbox{[O \textsc{iii}]}}

\newcommand{\NII}{\mbox{[N \textsc{ii}]}}
\newcommand{\HeI}{\mbox{He \textsc{i}}}
\newcommand{\Hb}{H$\beta$}
\newcommand{\Ha}{H$\alpha$}
\newcommand{\kms}{\mbox{km s$^{-1}$}}
\newcommand{\NHI}{\mbox{N(H \textsc{i})}}

\title[AGN-Driven Neutral Outflows at $z\sim$~2]{JWST Reveals Widespread AGN-Driven Neutral Gas Outflows in Massive $z\sim$~2 Galaxies}

\author[R. L. Davies et al.]{\href{https://orcid.org/0000-0002-3324-4824}{Rebecca L. Davies}$^{1,2}$\thanks{Contact e-mail: \href{mailto:rdavies@swin.edu.au}{rdavies@swin.edu.au}},
Sirio Belli,$^{3}$
Minjung Park,$^{4}$
J. Trevor Mendel,$^{5,2}$
Benjamin D. Johnson,$^{4}$
\newauthor Charlie Conroy,$^4$
Chloë Benton,$^6$
Letizia Bugiani,$^3$
Razieh Emami,$^4$
Joel Leja,$^{7,8,9}$
Yijia Li,$^{7,8}$
\newauthor Gabriel Maheson,$^{10,11}$
Elijah P. Mathews,$^{7,8,9}$
Rohan P. Naidu,$^{12}$
Erica J. Nelson,$^{6}$
Sandro Tacchella,$^{10,11}$
\newauthor Bryan A. Terrazas,$^{13}$
Rainer Weinberger$^{14}$
\\
$^1$Centre for Astrophysics and Supercomputing, Swinburne University of Technology, Hawthorn, Victoria 3122, Australia \\
$^2$ARC Centre of Excellence for All Sky Astrophysics in 3 Dimensions (ASTRO 3D), Australia \\
$^3$Dipartimento di Fisica e Astronomia, Università di Bologna, Bologna, Italy \\
$^4$Center for Astrophysics $\mid$ Harvard \& Smithsonian, Cambridge, MA, USA \\
$^5$Research School of Astronomy and Astrophysics, Australian National University, Canberra, ACT, Australia \\
$^6$Department for Astrophysical and Planetary Science, University of Colorado, Boulder, CO, USA \\
$^7$Department of Astronomy \& Astrophysics, The Pennsylvania State University, University Park, PA, USA \\
$^8$Institute for Gravitation and the Cosmos, The Pennsylvania State University, University Park, PA, USA \\
$^9$Institute for Computational \& Data Sciences, The Pennsylvania State University, University Park, PA, USA \\
$^{10}$Kavli Institute for Cosmology, University of Cambridge, Cambridge, UK \\
$^{11}$Cavendish Laboratory, University of Cambridge, Cambridge, UK \\
$^{12}$MIT Kavli Institute for Astrophysics and Space Research, Cambridge, MA, USA \\
$^{13}$Columbia Astrophysics Laboratory, Columbia University, New York, NY, USA \\
$^{14}$Leibniz Institute for Astrophysics, Potsdam, Germany
}

\pubyear{2024}

\begin{document}
\label{firstpage}
\pagerange{\pageref{firstpage}--\pageref{lastpage}}
\maketitle

\begin{abstract}
We use deep JWST/NIRSpec $R\sim$~1000 slit spectra of 113 galaxies at \mbox{1.7 $< z <$ 3.5}, selected from the mass-complete Blue Jay survey, to investigate the prevalence and typical properties of neutral gas outflows at cosmic noon. We detect excess \NaD\ absorption (beyond the stellar contribution) in 46\% of massive galaxies \mbox{($\log$ M$_*$/M$_\odot >$~10)}, with similar incidence rates in star-forming and quenching systems. Half of the absorption profiles are blueshifted by at least 100~\kms, providing unambiguous evidence for neutral gas outflows. Galaxies with strong \NaD\ absorption are distinguished by enhanced emission line ratios consistent with AGN ionization. We conservatively measure mass outflow rates of 3~--~100~$M_\odot$~yr$^{-1}$; comparable to or exceeding ionized gas outflow rates measured for galaxies at similar stellar mass and redshift. The outflows from the quenching systems \mbox{(log(sSFR)[yr$^{-1}$] $\lesssim$ -10}) have mass loading factors of \mbox{4~--~360}, and the energy and momentum outflow rates exceed the expected injection rates from supernova explosions, suggesting that these galaxies could possibly be caught in a rapid blowout phase powered by the AGN. Our findings suggest that AGN-driven ejection of cold gas may be a dominant mechanism for fast quenching of star formation at $z\sim$~2.
\end{abstract}

\begin{keywords}
galaxies: evolution -- galaxies: star formation -- galaxies: nuclei
\end{keywords}



\section{Introduction} \label{sec:intro}
Determining the physical mechanism(s) responsible for quenching star-formation in massive galaxies is key to our understanding of galaxy evolution. Cosmological simulations typically quench massive galaxies via feedback from active galactic nuclei (AGN) which both expels cold gas from galaxies and heats halo gas, preventing it from cooling and being re-accreted to replenish the reservoir of fuel for star-formation \citep[e.g.][]{DiMatteo05, Springel05, Bower06, Croton06, Hopkins06, Somerville08, Erb15, Beckmann17}. However, definitive observational evidence for a link between AGN feedback and star-formation quenching has yet to be established (see \citealt{Harrison17} and references therein). 

Over the past decade, large galaxy surveys have significantly improved our understanding of outflows during the peak epoch of star-formation and black-hole growth at $z\sim$~1~--~3, when feedback is expected to be most active. It is now well established that outflows are ubiquitous in massive star-forming galaxies at this epoch \citep[e.g.][]{Shapley03, Weiner09, Rubin14, Harrison16, NMFS19}. The strongest outflows are powerful enough to rapidly suppress star-formation in their host galaxies \citep[e.g.][]{CanoDiaz12, Cresci15, Carniani16, Kakkad16, Davies20}, but these are generally associated with the most luminous AGN which are present in a small fraction of massive galaxies at any given time. It remains unclear whether outflows driven by more typical AGN are capable of quenching star-formation in their host galaxies. Measurements based on optical emission lines (tracing ionized gas) suggest that most outflows remove gas less rapidly than it is consumed by star-formation \citep[e.g.][]{Harrison16, Leung19, NMFS19}, whilst UV absorption line measurements (tracing neutral gas) suggest that the mass outflow rates are comparable to the star-formation rates of the host galaxies \citep[e.g.][]{Weiner09, Kornei12}. 

Observations based on a single gas phase provide a very incomplete picture of outflows which contain gas at a range of temperatures and densities including hot (10$^{6-7}$~K) X-ray emitting gas, warm (10$^{4-5}$~K) ionized gas, cool (100~K) neutral gas and cold (10~K) molecular gas \citep[e.g.][]{Strickland04, Feruglio10, Leroy15, Krieger19}. Despite significant observational advances, the vast majority of outflows have only been observed in one gas phase, and constraining the total mass of gas ejected by outflows remains very challenging. State-of-the-art simulations of star-formation driven outflows predict that the majority of the outflowing mass is carried in the neutral and molecular phases \citep[e.g][]{Kim20} (although it remains unclear whether cool clouds survive to large galactocentric radii or are shredded by the hot wind; e.g. \citealt{Schneider20, Fielding22}). Outflowing molecular gas is notoriously difficult to detect \citep[see][and references therein]{Veilleux20} but is often found to carry much more mass than the ionized phase \citep[e.g.][]{Fluetsch19, Vayner17, Brusa18, HerreraCamus19}. Cool neutral gas in outflows is commonly probed using low ionization rest-frame far-UV absorption lines. However, it is difficult to detect the far-UV continuum of massive, dusty AGN host galaxies, and UV absorption line measurements at high redshift are generally restricted to the strongest transitions which are often saturated, providing only lower limits on the outflowing mass \citep[see][and references therein]{Veilleux20}.

An alternative tracer of neutral outflows is the resonant \NaD~$\lambda \lambda$~5891,5897\AA\ doublet. With a first ionization potential of 5.1~eV, Na~\textsc{i} exists primarily in neutral regions where it is shielded by significant columns of gas and dust \citep[e.g.][]{Savage96, Baron20}. Due to its location in the rest-frame optical spectrum, observations of \NaD\ already exist for many thousands of nearby galaxies. A few percent of local massive star-forming galaxies show blueshifted \NaD\ absorption indicative of neutral gas outflows \citep[e.g.][]{Nedelchev19, Avery22}. In galaxies with both neutral and ionized outflows, the neutral outflow rates are 10~--~100 times larger \citep[e.g.][]{RobertsBorsani20, Avery22, Baron22}, confirming that ionized gas likely represents a small fraction of the total mass budgets of typical nearby outflows. The \NaD\ absorption originates on spatial scales $\lesssim$~10~kpc \citep[e.g.][]{Martin06, Rupke15, Rupke17, Baron20, RobertsBorsani20, Avery22, Rubin22}, indicating that it traces recently launched outflows rather than gas in the circumgalactic medium. Although \NaD\ is easily accessible in the local Universe, it has been significantly harder to detect at $z\sim$~1~--~3 where the line shifts into the observed near-infrared. 

The unprecedented infrared sensitivity of JWST enables the detection of \NaD\ in distant galaxies, providing a new probe of neutral outflows in the early Universe. Initial observations have already revealed \NaD\ absorption tracing neutral outflows in three AGN host galaxies at \mbox{$z\sim$~2~--~3}, of which one is a quasar \citep{Cresci23, Veilleux23} and two are post-starburst galaxies \citep{Belli23, DEugenio23}. These outflows are also detected in ionized gas emission lines, enabling direct comparisons of the mass outflow rates in different gas phases. Focusing on the post-starburst galaxies, both \citet{Belli23} and \citet{DEugenio23} find that the neutral mass outflow rates are about two orders of magnitude larger than the ionized outflow rates and exceed the current star formation rates (SFRs) of the host galaxies. The rapid ejection of cold gas by powerful AGN-driven outflows may have led to the recent, fast quenching of star-formation in these galaxies. 

The detection of strong neutral gas outflows in two post-starburst AGN host galaxies provides tantalizing evidence that ejective AGN feedback may be an important mechanism for quenching massive galaxies at cosmic noon. However, it is unclear whether these objects are representative of the overall galaxy population. In this paper, we characterize the incidence and typical properties of neutral outflows across the galaxy population using 113 galaxies at \mbox{1.7 $< z <$ 3.5} from the mass-selected Blue Jay survey. We discuss the sample and observations in Section \ref{sec:data}, present the census of \NaD\ absorption in Section \ref{sec:census} and examine the neutral outflow properties in Section \ref{sec:outflow_properties}. We discuss the connection between neutral outflows, AGN activity and star-formation quenching in Section \ref{sec:discussion} and present our conclusions in Section \ref{sec:conclusion}.

\section{Observations and Data reduction}\label{sec:data}
\subsection{Blue Jay}
This work is based on observations from the JWST Cycle 1 program Blue Jay (GO 1810; PI Belli). The NIRSpec micro-shutter assembly (MSA; \citealt{Ferruit22, Rawle22}) was used to obtain R~$\simeq$~1000 spectra of 151 galaxies spread over two masks in the COSMOS field. Four of these galaxies are filler targets at $z\sim$~6, and the remaining 147 galaxies form a mass-selected sample (\mbox{9 $< \log(M_*/M_\odot) <$ 11.5}) at cosmic noon (\mbox{1.7 $< z <$ 3.5}). All galaxies were observed using the three medium resolution gratings (G140M, G235M and G395M) with exposure times of 13h, 3.2h and 1.6h respectively. A slitlet made of at least 2 MSA shutters was placed on each target and we employed a 2-point A-B nodding pattern along the slit. The data were reduced using a modified version of the JWST Science Calibration Pipeline v1.10.1, and version 1093 of the Calibration Reference Data System. Master background subtraction was performed using a spectrum measured from dedicated background slits and galaxy 1D spectra were optimally extracted \citep{Horne86}. The spectrum extraction failed for 6 galaxies which are excluded from our sample. Full details of the Blue Jay sample selection, observations and data reduction will be provided in the survey paper (Belli et al., in prep).

\begin{figure*}
\centering
 \includegraphics[scale=1.0, clip = True, trim = 0 12 0 0]{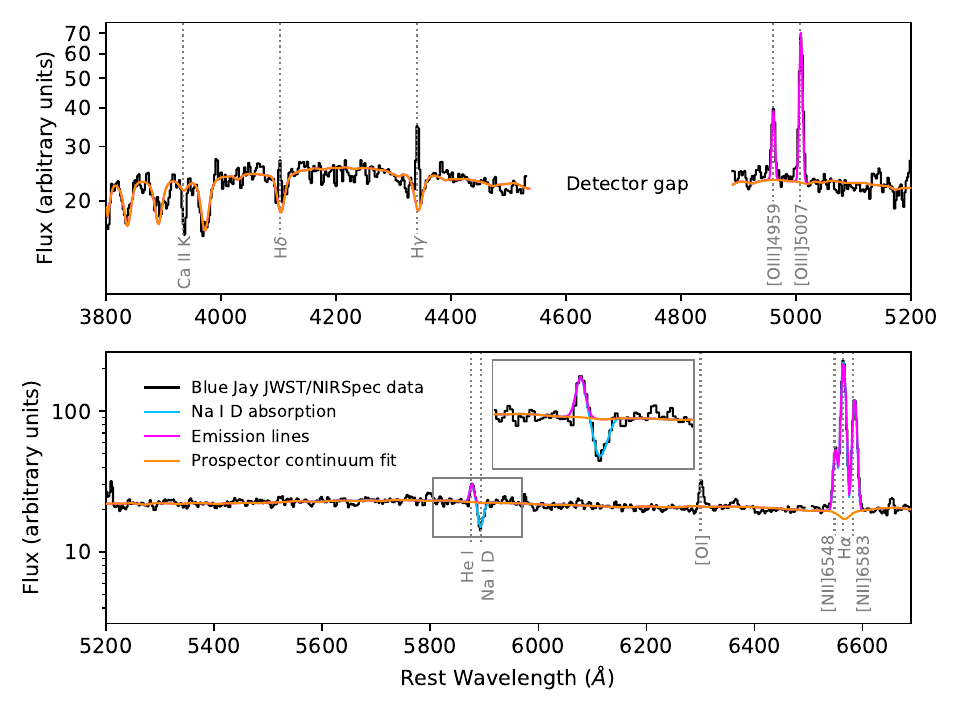}
\caption{R~$\simeq$~1000 NIRSpec MSA spectrum of COSMOS-10245 at $z$ = 1.81 over rest-frame wavelengths of 3800~--~6700\AA\ (black). The orange curve shows the best-fit stellar continuum model from \textsc{Prospector} and the magenta and blue curves show the best-fit models for the strong emission lines and the \NaD\ doublet absorption, respectively, obtained from MCMC fitting. The inset in the bottom row shows a zoom-in on the region around the He~\textsc{I} and \NaD\ lines. The observed \NaD\ absorption is significantly stronger than expected for the stellar component.} \label{fig:summary_spec} 
\end{figure*}

The individual grating spectra were combined to produce wide spectra covering rest-frame wavelengths of at least 3000\AA~--~1.2$\mu$m for all galaxies. Figure \ref{fig:summary_spec} shows the spectrum of COSMOS-10245 at $z$~=~1.81 over rest-frame wavelengths of \mbox{3800~--~6700\AA}. The wavelength coverage is not continuous due to gaps between the NIRSpec detectors, and we excluded 7 galaxies for which \NaD\ falls within a detector gap. Finally, we excluded 21 galaxies for which no spectroscopic redshift could be determined due to an absence of identifiable emission or absorption line features. Our final sample consists of 113 galaxies and includes COSMOS-11142, the post-starburst galaxy analysed by \citet{Belli23}.

\subsection{Stellar Population Fitting}
The goal of this paper is to search for neutral gas outflows traced by interstellar \NaD\ absorption. However, \NaD\ absorption can also originate in stellar atmospheres and is particularly prominent in late-type stars \citep[e.g.][]{OConnell76, Peterson76, Carter86, Alloin89, Worthey98}. Therefore, it is imperative to accurately remove the stellar absorption contribution prior to our analysis. 

We model the stellar continuum using \textsc{Prospector}, a Bayesian stellar population inference code designed to simultaneously fit photometry and spectroscopy spanning UV to mid-IR wavelengths \citep{Johnson21}. We adopt the synthetic stellar population library \textsc{fsps} \citep{Conroy09, Conroy10}, the \textsc{mist} isochrones \citep{Choi16}, and the Chabrier initial mass function. The stellar metallicity is free to vary and individual elemental abundances are assumed to be solar scaled. We note that varying the stellar [Na/Fe] ratio within a reasonable range does not significantly impact our results (see Section \ref{subsec:incidence}). We adopt a non-parametric star-formation history with 14 bins spaced logarithmically in time except for the lowest age bin which is placed at \mbox{30 Myr}. \textsc{Prospector} accounts for dust absorption and re-emission which are assumed to be in energy balance. The dust absorption model consists of a primary component which applies to all stars and follows the \citet{Kriek13} attenuation curve, as well as a multiplicative term representing extra attenuation towards young stars (with ages \mbox{$<$~10~Myr}). The \textsc{Prospector} model also includes a multiplicative `jitter' term that scales the measurement errors to better represent the statistical fluctuations in the data, as well as a polynomial distortion term that corrects for shape mismatches between the spectra and the stellar templates resulting from imperfect flux calibration and/or slit losses. 

We use \textsc{Prospector} to fit the JWST spectra along with publicly available HST/ACS+WFC3 \citep{Skelton14, Momcheva16} and Spitzer/IRAC \citep{Laigle16} photometry. The NIRSpec observations cover many age-sensitive spectral features including the 4000\AA\ break and the Balmer absorption series (see Figure \ref{fig:summary_spec}), providing strong constraints on the stellar population properties. During the fitting, we mask prominent emission lines as well as the \NaD\ and Ca~\textsc{ii h + k} absorption lines which can have significant contributions from interstellar gas. Full details of the \textsc{Prospector} fitting will be provided in Park et al. (in prep). 

The orange curve in Figure \ref{fig:summary_spec} shows the best-fit stellar continuum model for COSMOS-10245. The model provides a very good fit to the well-detected Balmer absorption series and enables us to accurately quantify the stellar contribution to the observed \NaD\ absorption. In this case, the observed \NaD\ absorption is significantly stronger than expected from the stellar continuum alone. Unless otherwise noted, all subsequent references to \NaD\ absorption refer to absorption in excess of the stellar contribution. 

\textsc{Prospector} outputs probability distribution functions which are used to calculate the best-fit values (median) and corresponding uncertainties (16th-84th percentile range) for all model parameters. These include the distortion polynomial (which we use to produce flux-calibrated spectra), the best-fit jitter term (used to scale the measurement errors), the stellar mass (M$_*$) and the non-parametric star-formation history. The SFRs reported in this paper refer to the SFR in the youngest age bin (averaged over the last 30 Myr), but similar results are obtained using SFRs averaged over 100 Myr or SFRs computed from the \Ha\ emission line luminosity.

\subsection{Emission and absorption line fitting}\label{subsec:combined_fitting}
We fit the spectrum of each Blue Jay galaxy over the wavelength region between 3800~--~6700\AA, including contributions from the stellar continuum ($F_{\rm *, Prospector}$), ionized gas emission lines ($F_{\rm gas}$) and excess \NaD\ absorption ($F_{\rm Na D, excess}$):
\begin{equation}\label{eqn:spec_model}
F(v) = \left[ F_{\rm *, Prospector} + F_{\rm gas} \right] \times F_{\rm Na D, excess}
\end{equation}
The emission and absorption line components must be fit simultaneously because the \HeI~$\lambda$~5876\AA\ emission line falls in close proximity to \NaD\ (see Figure \ref{fig:summary_spec}). The emission and absorption line models are convolved with the wavelength-dependent NIRSpec line spread function prior to fitting\footnote{We adopt the nominal resolution for uniform slit illumination  from JDox, but note that the true resolution is notably higher than this for compact sources \citep[e.g.][]{DeGraaff23}. As a consequence, the measured velocity dispersions represent lower limits on their true values. \label{footnote:spectral_res}}.

The emission line model includes the two strongest Balmer lines (\Ha\ and \Hb), the two strongest forbidden line doublets within the fitted wavelength region (\OIII~$\lambda \lambda$~4959,5007\AA\ and \NII~$\lambda \lambda$~6548,6583\AA), and \HeI~$\lambda$~5876\AA. Outflowing gas can produce redshifted \NaD\ emission due to resonant scattering off gas in the receding side of the outflow \citep[e.g.][]{Prochaska11}, but this has only been observed in a handful of objects, \citep[e.g.][]{Rupke15, Perna19, Baron20, Baron22, Sun23}, and we do not find clear evidence for \NaD\ emission in any of our spectra. 

We initially fit each emission line with a single Gaussian profile, constraining all lines to have the same velocity offset and dispersion. A single Gaussian component is sufficient to explain the vast majority of observed line profiles given the relatively low spectral resolution of the observations. Two galaxies show complex Balmer and forbidden line profiles that are not well represented by a single kinematic component, and for these objects we add an extra Gaussian component to all 7 emission lines. Two other galaxies show prominent AGN broad line region emission which is modelled as a broad Gaussian component in the \Ha, \Hb\ and \HeI\ lines. 

Interstellar \NaD\ absorption is parametrized using the standard partial covering model \citep{Rupke05a}:
\begin{equation}\label{eqn:partial_covering}
F_{\rm Na D, excess}(v) = 1 - C_f + C_f \textrm{exp} \left( - \tau_b(v) - \tau_r(v) \right)
\end{equation}
Here, $C_f$ is the covering fraction of the absorbing gas against the background continuum source, and $\tau_b(v)$ and $\tau_r(v)$ are the optical depth profiles of the blue (\NaD~$\lambda$~5891\AA) and red (\NaD~$\lambda$~5897\AA) doublet lines, respectively. We assume that the optical depth has a Gaussian velocity distribution:
\begin{equation}
\tau(v,\sigma) = \tau_0 \, \textrm{exp}(-v^2/2\sigma^2)
\end{equation}
The optical depth at the centre of the blue line ($\tau_{0, b}$) is fixed to be twice the optical depth at the centre of the red line ($\tau_{0, r}$), reflecting the known doublet ratio.\footnote{The equivalent width ratio is not fixed and varies between 2 in the optically thin regime and 1 in the optically thick regime. This is because the curve of growth representing the relationship between optical depth and equivalent width is non-linear.} We fit each \NaD\ line using a single Gaussian velocity distribution, which is sufficient to describe the observed absorption profiles in all cases. The spectral resolution is comparable to the velocity separation between the doublet lines, making it very difficult to resolve multi-component velocity structure. The \NaD\ absorption kinematics are allowed to vary independently of the emission line kinematics. 

The absorption optical depth and covering fraction can become degenerate when the \NaD\ doublet is blended (e.g. \citealt{Rupke05a}; see discussion in Appendix \ref{sec:appendix}). To obtain accurate constraints on the parameter uncertainties and degeneracies, we perform the fitting using \textsc{emcee} \citep{ForemanMackey13}, an Affine Invariant Markov Chain Monte Carlo (MCMC) Ensemble sampler. The walkers are initialised in small regions around the best-fit values obtained from preliminary least squares fitting. 

Similarly to the \textsc{Prospector} parameters, the best-fit emission and absorption line parameters represent the medians of the \textsc{emcee} posterior distributions and the error bars reflect the 16th~--~84th percentile ranges. The best-fit emission line and \NaD\ absorption line profiles for COSMOS-10245 are shown by the magenta and blue curves in Figure \ref{fig:summary_spec}, respectively. 

\begin{figure*}
\centering
 \includegraphics[scale=0.8]{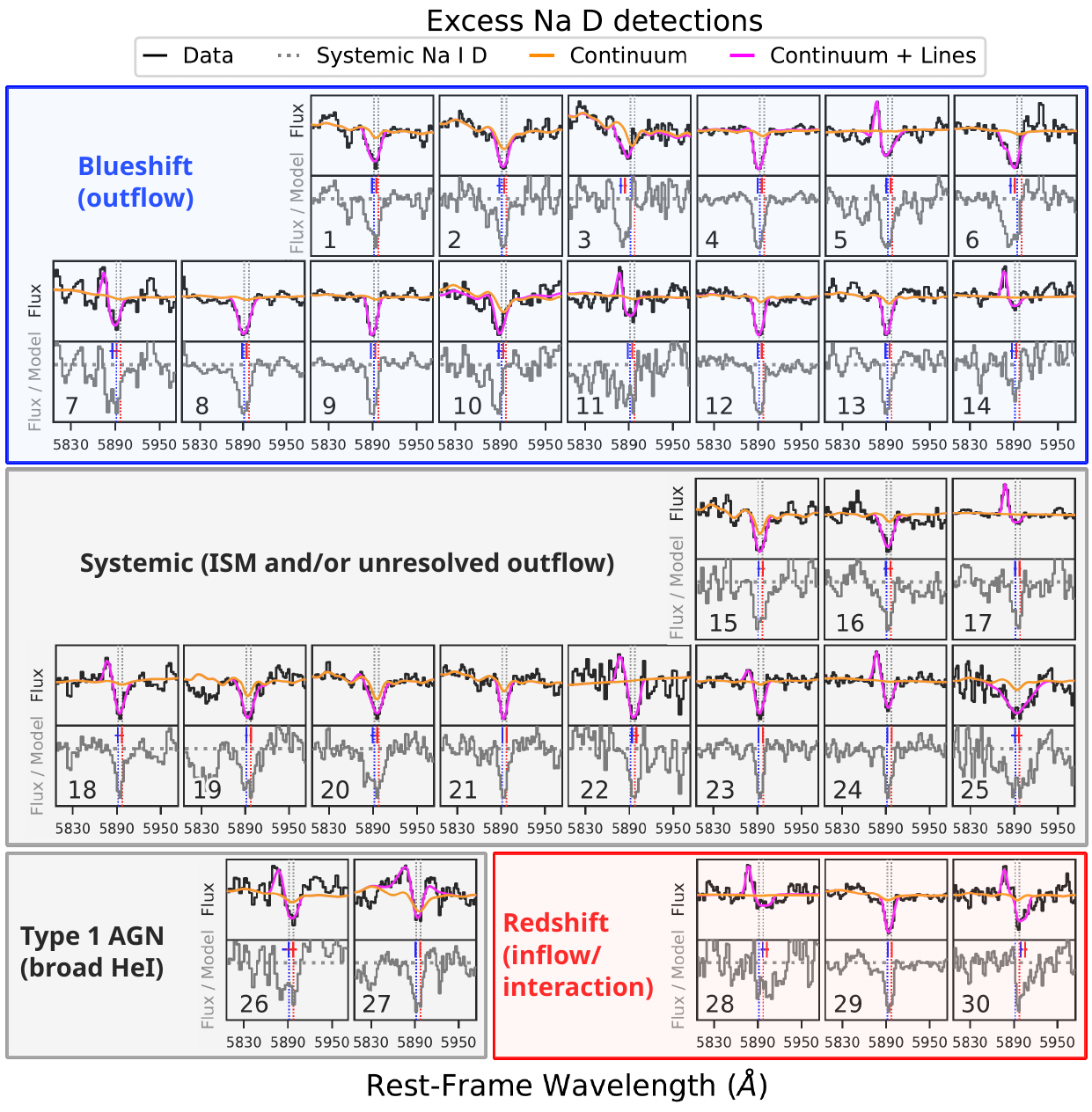}
\caption{Spectral cutouts covering 5805~--~5975\AA\ for the 30 galaxies with \NaD\ absorption significantly exceeding the stellar contribution. Each pair of panels represents an individual galaxy and shows the observed spectrum (top, black), the best-fit continuum-only (orange) and continuum + line (magenta) models, and the residual spectrum after removing the continuum and line emission (bottom, grey). Numbers in the lower panels match the sample IDs in Table \ref{table:absorption_properties}. The $y$-axes are scaled independently for each panel. The dotted vertical lines show the systemic wavelengths of the blue and red \NaD\ lines in the galaxy rest frames, and the blue and red symbols in the lower (residual) panels show the centroids of the best-fit line profiles (centered at the 50th percentile velocity with horizontal error bars indicating the 16th~--~84th percentile confidence intervals). The absorption profiles are grouped into four categories: blueshifted (top), systemic (middle), Type 1 AGN with broad \HeI\ emission (bottom left) and redshifted (bottom right). Some absorption profiles appear shifted in velocity but are classified as systemic because the 68\% confidence interval encompasses zero velocity.} \label{fig:NaD_detections} 
\end{figure*}

\section{A census of \NaD\ absorption at $z\sim$~2}\label{sec:census}
\subsection{Incidence}\label{subsec:incidence}
We visually inspect the fits for all Blue Jay galaxies and identify sources with significant interstellar \NaD\ absorption. For the first time, we report evidence for widespread \NaD\ absorption in $z\sim$~2 galaxies, as shown in Figure \ref{fig:NaD_detections}. The absorption profiles are grouped into four categories based on the velocity shift of the \NaD\ absortion feature (see Section \ref{subsec:absorption_origin}). Within each category, each pair of panels represents an individual galaxy and shows the observed spectrum over the region covering $\pm$~85\AA\ around the \NaD\ doublet (top, black), the best-fit continuum-only (orange) and continuum + line (magenta) models, and the residual spectrum after removing the continuum and line emission (bottom, grey). From our initial sample of 113 galaxies, 30 galaxies (27\%) have \NaD\ absorption much stronger than expected from the stellar populations alone. The MCMC posterior distributions confirm that the excess absorption is detected at $\geq$~3$\sigma$ significance in all cases. 

\begin{figure*}
\centering
\includegraphics[scale=0.95]{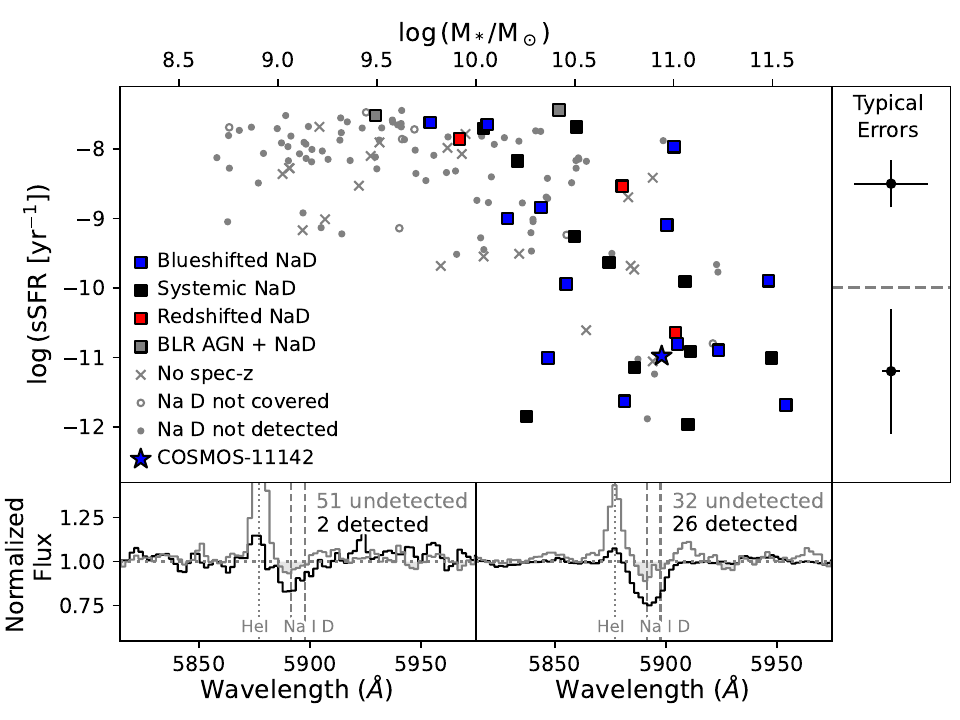}
\caption{Top: Distribution of galaxies in the $M_*$--~sSFR plane. Galaxies with detected \NaD\ absorption are shown as squares, where the colors reflect the velocity shift of the absorption as shown in Figure \ref{fig:NaD_detections}. Filled grey circles indicate galaxies with no significant absorption, open circles indicate galaxies for which \NaD\ falls in a detector gap and `x` symbols indicate galaxies with no spectroscopic redshift. We detect \NaD\ absorption in 46\% of massive galaxies ($\log M_*/M_\odot >$ 10), distributed almost uniformly over more than four orders of magnitude in sSFR. The panel on the right shows the typical $M_*$ and sSFR errors for galaxies in two sSFR bins divided at \mbox{log(sSFR[Gyr$^{-1}$]) = -10}. Bottom: Stacked spectra of galaxies with and without individual \NaD\ detections (black and grey, respectively) in two stellar mass bins divided at \mbox{log($M_*/M_\odot$) = 10}. The two \NaD-detected galaxies with Type 1 AGN are omitted from the stacks. Stacking does not reveal any additional absorption in low mass galaxies. There is some evidence for weak systemic absorption in high mass galaxies lacking individual \NaD\ detections.}\label{fig:main_sequence} 
\end{figure*}

The main panel of Figure \ref{fig:main_sequence} shows how the \NaD\ detections (colored squares) are distributed as a function of stellar mass ($M_*$) and specific SFR (sSFR). Galaxies without \NaD\ absorption are shown with filled grey circles. It is clear that interstellar \NaD\ absorption is detected almost exclusively in massive galaxies. 27/59 (46\%) of the $\log(M_*/M_\odot) >$~10 galaxies with spectroscopic redshift measurements show \NaD\ absorption, whereas the detection fraction in lower mass galaxies is 6\%. Interestingly, the detections are spread almost uniformly over four orders of magnitude in sSFR, from highly star-forming galaxies to quenching systems. The prevalence of interstellar \NaD\ absorption in the mass-selected Blue Jay sample indicates that large neutral gas reservoirs are prevalent in massive $z\sim$~2 galaxies.

The incidence of interstellar \NaD\ absorption is not strongly dependent on the assumed sodium abundance. We generate stellar absorption profiles for a range of sodium abundances using the \textsc{alf} code \citep{Conroy12, Conroy18}, and find that extremely sodium-enhanced stellar populations (\mbox{[Na/Fe] = +0.6}) can produce excess absorption with a rest-frame equivalent width of up to 1.2\AA. This is weaker than all our observed excess absorption profiles which have measured equivalent widths (after removing the stellar contribution) of 1.5~--~11.4~\AA\ (median 4.4\AA; see Table \ref{table:absorption_properties}).

\subsubsection{Stacking to search for weaker NaD absorption}
The detection of interstellar \NaD\ absorption requires a robust measurement of the stellar continuum, so our \NaD\ sample may be biased towards the brightest continuum sources. We search for interstellar absorption in galaxies lacking individual \NaD\ detections by stacking them in two bins of stellar mass (above and below $\log(M_*/M_\odot)$~=~10). The spectra are continuum-normalized prior to stacking and weighted by the average continuum signal-to-noise ratio within 150\AA\ of the \NaD\ line. Unweighted stacks are noisier but lead to the same overall conclusions.

\begin{figure*}
\centering
\includegraphics[scale=1, clip = True, trim = 0 100 0 0]{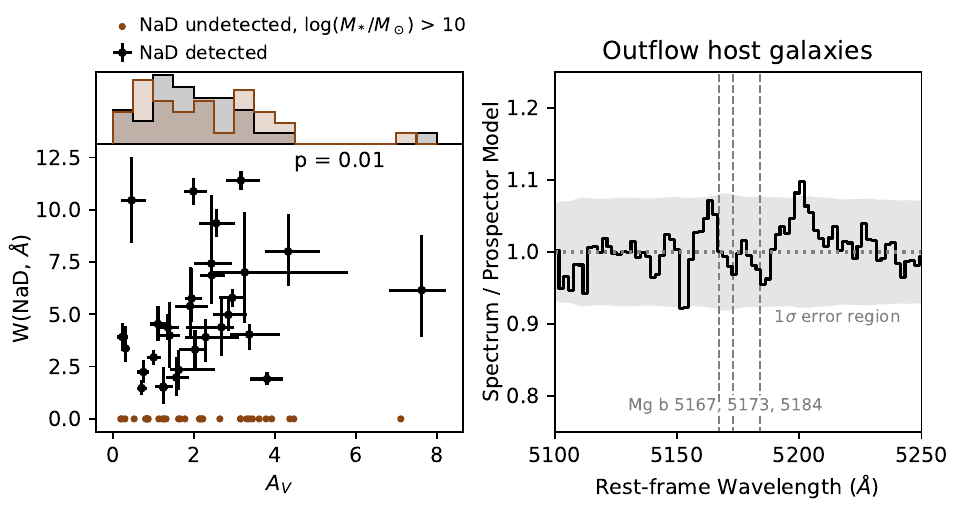}
\caption{Left: Rest-frame equivalent width of excess \NaD\ absorption as a function of $A_V$ for all \NaD-detected galaxies (black). The $A_V$ values include both the primary attenuation component and the extra attenaution towards young stars. The $A_V$ errors represent the 16th-84th percentile range from the \textsc{Prospector} posterior probability distributions. The \NaD\ absorption strength correlates with $A_V$, indicating that the absorption is interstellar in origin. Galaxies lacking \NaD\ detections are shown as brown circles with W(\NaD) artificially set to zero. The upper histograms show the $A_V$ distributions of massive galaxies with and without detected \NaD\ absorption. The two distributions do not differ significantly. Right: Stacked residual spectra of outflow host galaxies, zoomed-in on the \MgI\ triplet. There is no significant residual \MgI\ absorption, indicating that the stellar absorption is fully accounted for in the \textsc{Prospector} fitting.} \label{fig:low_ssfr_absorption}
\end{figure*}

The bottom panels of Figure \ref{fig:main_sequence} show stacked spectra of galaxies with and without individual \NaD\ detections (black and grey, respectively). Even after stacking 51 galaxies, we do not find any evidence of excess \NaD\ absorption in low mass galaxies lacking individual \NaD\ detections. There are at least two possible reasons for this. Firstly, Na~\textsc{I} has an ionization potential of 5.1~eV and therefore cannot exist in large quantities without a significant amount of dust shielding. In the local Universe there is a well-known relationship between \NaD\ absorption strength and dust attenuation (quantified by e.g. the $V$ band line-of-sight attenuation $A_V$, the color excess \mbox{E(B-V)}, or the Balmer decrement $f$(\Ha)/$f$(\Hb); e.g. \citealt{Heckman00, Veilleux05, Chen10, Veilleux20, Avery22}, and this correlation is also seen in our sample (see left-hand panel of Figure \ref{fig:low_ssfr_absorption}). Most low mass galaxies likely do not have enough dust and metals to retain detectable amounts of neutral sodium. A second possibility is that we do not see strong excess \NaD\ absorption because it primarily traces AGN-driven outflows (see Sections \ref{subsubsec:blueshift_origin} and \ref{sec:outflow_properties}), and these are rare in low mass galaxies \citep[e.g.][]{Genzel14, NMFS19, Leung19}.

There is some evidence for weak excess \NaD\ absorption in the stack of 32 massive galaxies lacking individual detections. This excess absorption falls at approximately zero velocity and could plausibly be explained by additional stellar absorption (see Section \ref{subsubsec:blueshift_origin}). Regardless, it is significantly weaker than the absorption seen in individually detected galaxies. The fact that we see strong interstellar \NaD\ absorption in 46\% of massive galaxies whilst the remaining 54\% show weak or no absorption suggests that the distribution of absorption strengths is not continuous, and may be more bimodal. 

\subsubsection{Link between strong \NaD\ absorption and galaxy properties}
The presence of strong \NaD\ absorption in \textit{massive} galaxies does not appear to be governed by dust properties: when restricting the sample to galaxies with \mbox{$\log(M_*/M_\odot) >$~10}, there is no significant difference between the $A_V$ or E(B-V) distributions galaxies with and without \NaD\ absorption (see histograms in the upper left panel of Figure \ref{fig:low_ssfr_absorption}). We investigate whether the \NaD\ absorption strength may instead be related to galaxy inclination. Detections of strong excess \NaD\ absorption in the local Universe seem to be preferentially associated with outflows \citep[e.g.][]{Heckman00, Rupke05a, Cazzoli16, Rupke17}. Blueshifted wind material is primarily observed in face-on galaxies \citep[e.g.][]{Chen10, RobertsBorsani19, Avery22} and the magnitude of the line-of-sight velocity shift increases with decreasing inclination \citep[e.g.][]{Bae18, Concas19, Sun23}, suggesting that the neutral outflows are launched perpendicular to the galaxy disk. In the Blue Jay sample, we do not find any clear relationship between the presence of \NaD\ absorption and the galaxy axis ratio measured from CANDELS HST imaging \citep{vanderWel12}. However, this imaging covers rest-frame UV and blue-optical wavelengths where young stellar populations dominate. Ongoing JWST NIRCam and MIRI imaging of the COSMOS field will enable more complete mapping of the galaxy stellar mass distributions and thus more accurate axis ratio measurements in the future. 

Intriguingly, the most striking difference between the galaxies with and without \NaD\ absorption is their \NII/\Ha\ ratios, which are discussed further in Section \ref{subsec:line_ratios}. 

\subsection{Origin of absorbing gas}\label{subsec:absorption_origin}
The origin of the interstellar \NaD\ absorption can be determined by examining the velocity of the absorption relative to the galaxy systemic velocity (shown by the vertical dotted grey lines in Figure \ref{fig:NaD_detections}). In many cases, the excess \NaD\ absorption is clearly offset from the galaxy systemic redshift, with measured velocity shifts ranging from \mbox{-680} to +400 \kms. We use the absorption velocity posterior probability distributions to classify the galaxies into three categories: blueshifted absorption (84th percentile velocity less than zero), redshifted absorption (16th percentile velocity greater than zero), and systemic absorption (consistent with zero). These classifications are shown in Figure \ref{fig:NaD_detections} and determine the colors of the markers in Figure \ref{fig:main_sequence}. The absorption velocities for the two broad line AGN are very sensitive to the fitting of the broad \HeI\ emission, so we cannot reliably classify them. Of the remaining 28 galaxies, 14 (50\%) show blueshifted absorption (discussed in Section \ref{subsubsec:blueshift_origin}), 11 (39\%) are consistent with the systemic velocity (Section \ref{subsubsec:systemic_origin}), and 3 (11\%) show redshifted absorption (Section \ref{subsubsec:redshift_origin}).

\subsubsection{Outflowing gas}\label{subsubsec:blueshift_origin}
Half of the classifiable absorption profiles (14/28 or 50\%) are blueshifted by at least 100~\kms\ (50th percentile velocity), which is an unambiguous sign of neutral gas outflows \citep[e.g.][]{Phillips93, Heckman00, Rupke02, Schwartz04, Martin05, Rupke05a}. The high fraction of outflows among the \NaD-detected galaxies is consistent with studies of local (ultra-)luminous infrared galaxies which find that excess \NaD\ absorption is preferentially blueshifted and associated with winds \citep[e.g.][]{Heckman00, Rupke05a, Cazzoli16, Rupke17}. The overall incidence of \NaD\ outflows across the full mass range of our sample \mbox{($\log(M_*/M_\odot)$ = 8.5 -- 11.5)} is 12\% (14/113). This is notably higher than the $\sim$~1\% incidence of neutral outflows among typical star-forming and AGN host galaxies with similar stellar masses at $z\sim$~0 \citep[e.g.][]{Nedelchev19, Avery22}, but comparable to the $\sim$~20\% incidence of outflows in local post-starburst galaxies \citep[e.g.][]{Sun23}. Compared to other measurements at $z\sim$~2, the observed frequency of \NaD\ outflows is lower than the \mbox{20~--~30\%} incidence of ionized outflows traced by optical emission lines among massive galaxies \citep[e.g.][]{NMFS19}, and significantly lower than the $\sim$~100\% incidence of neutral outflows traced by rest-frame UV absorption lines in UV-bright galaxies \citep[e.g.][]{Steidel10}. We note that the \NaD\ outflow fraction may be underestimated by up to a factor of 2 at high stellar masses due to the low spectral resolution of the observations (R~$\sim$~1000). Galaxies with strong systemic absorption could have weaker outflow components that would not be detectable in our data (see Section \ref{subsubsec:systemic_origin}). The low incidence of \NaD\ outflows in low mass galaxies is likely driven to a large degree by the requirement for dust shielding which prevents the detection of \NaD\ absorption in UV-bright galaxies \citep[see also][]{Avery22}. Rest-frame UV and \NaD\ absorption lines may probe outflows in almost entirely separate populations of galaxies. 

The incidence of neutral outflows in our sample appears to be independent of star-formation activity. The outflow sources, indicated by the blue squares in Figure \ref{fig:main_sequence}, are distributed over a wide range in sSFRs, extending all the way to the quenching galaxy regime. This is somewhat surprising given that in the local Universe, blueshifted \NaD\ absorption is preferentially found in highly star-forming systems \citep[e.g.][]{Rupke05c, Chen10, Concas19, RobertsBorsani19, Avery22}. Excess \NaD\ absorption has been found in many nearby massive quiescent galaxies \citep[e.g.][]{Carter86, Alloin89, Worthey98, Thomas03, Worthey11, Concas19, RobertsBorsani19}, but this absorption is typically consistent with the systemic velocity. Furthermore, the \NaD\ absorption strength in local early type galaxies is found to correlate with that of the \MgI\ triplet, which only arises in stellar atmospheres \citep[e.g.][]{Heckman00, Rupke02, Rupke05a, Alatalo16}. Therefore, the excess \NaD\ absorption is generally attributed to enhanced stellar absorption, perhaps due to elevated [Na/Fe] \citep[e.g.][]{OConnell76, Peterson76, Parikh18} and/or a bottom-heavy initial mass function \citep[e.g.][]{vanDokkum10, vanDokkum12, Spiniello12}. 

The excess \NaD\ absorption we detect in low sSFR galaxies at $z\sim$~2 has distinctly different properties from the excess stellar absorption seen at $z\sim$~0. Firstly, all sources classified as outflows have \NaD\ absorption blueshifted by at least 100~\kms, indicating that a non-systemic component is required to explain the observed absorption profile. Secondly, the observed \NaD\ absorption is too strong to explain with excess stellar absorption (see Section \ref{subsec:incidence}). \citet{Jafariyazani20} similarly detected excess \NaD\ absorption in a lensed quiescent galaxy at $z\sim$~2 and showed that it was too strong to be explained by enhanced Na abundance. Thirdly, the excess \NaD\ absorption is not associated with excess \MgI\ absorption. The right-hand panel of Figure \ref{fig:low_ssfr_absorption} shows stacked residual spectra of galaxies with neutral gas outflows, zoomed in on the region around the \MgI\ line. There is no evidence for significant excess absorption, suggesting that the stellar absorption contribution has been fully accounted for in the \textsc{Prospector} fitting. Furthermore, the outflow host galaxies in our sample fall above the \NaD~--~\MgI\ correlation observed in local early-type galaxies \citep{Alatalo16}, indicating that the \NaD\ absorption is much stronger than expected based on the \MgI\ absorption. Finally, the \NaD\ absorption strength (quantified by the rest-frame equivalent width $W({\rm Na D})$) is positively correlated with $A_V$ (Figure \ref{fig:low_ssfr_absorption}, left), indicating that the excess \NaD\ absorption is interstellar in origin. The combination of these factors provides strong evidence that the blueshifted \NaD\ absorption observed in massive, low sSFR galaxies at $z\sim$~2 traces neutral gas outflows.

\subsubsection{Systemic absorption}\label{subsubsec:systemic_origin}
A further 39\% (11/28) of the classified \NaD\ absorption profiles have centroid velocities consistent with the galaxy systemic velocity. This means that most of the neutral gas follows the bulk motion of the galaxy i.e. it is likely to be located primarily in the interstellar medium (ISM). Massive, high redshift galaxies are known to harbour large cold gas reservoirs \citep[see][and references therein]{Tacconi20}, and these could be responsible for producing the strong systemic absorption we observe. 

However, it is also possible that a non-negligible fraction of the absorption classified as systemic could arise from outflows. Firstly, we are only able to robustly identify outflows with velocity offsets exceeding $\sim$~100~\kms\ due to the relatively low spectral resolution of the observations and the close relative proximity of the \NaD\ doublet lines. Secondly, our classification based on the average absorption velocity would not identify outflow components that are hidden underneath strong ISM absorption. Observations at higher spectral resolution (i.e. $R$~=~2700 with JWST/NIRSpec) would enable us to perform 2-component fitting and separate neutral gas in the ISM from outflowing material (see Section \ref{subsec:mdot_calc}).

\subsubsection{Infalling gas}\label{subsubsec:redshift_origin}
The remaining 11\% (3/28) of the classified \NaD\ absorption profiles are redshifted, with velocity offsets of 27, 255 and 404~\kms. We note that although the absorption towards COSMOS-19572 is formally classified as infalling, the absolute velocity offset of 27~\kms\ is small and comparable to the measurement error, and therefore this absorption could plausibly be systemic. The other two redshifted absorption sources have significantly larger velocity shifts, suggesting that there is neutral gas flowing \textit{towards} these galaxies. The infalling material could originate in bulk flows within interacting systems or could be directly accreting onto the galaxies, providing cold gas that may sustain (or rejuvenate) star-formation. 

Redshifted \NaD\ absorption has been observed in local galaxies. \citet{RobertsBorsani19} find that redshifted absorption is prevalent in massive edge-on star-forming galaxies, consistent with a picture where the absorption traces gas accreting along the disk plane, potentially originating from galactic fountains. \citet{Roy21} find evidence for neutral gas inflows in a substantial fraction of passive `red geyser' galaxies, again likely originating from internal recycling and/or minor mergers (see also \citealt{Cheung16}).

We investigate the likely origin of redshifted absorption in the Blue Jay galaxies by examining their star formation histories and morphologies (using HST/ACS+WFC3 imaging from the 3D-HST survey and JWST/NIRCam imaging from the PRIMER survey; GO 1837, PI Dunlop). COSMOS-19572 shows evidence for a nearby companion and could be an interacting system. This galaxy also shows strong N~\textsc{ii} and S~\textsc{ii} line emission which could plausibly trace shocks induced by tidal forces. High spatial resolution maps of emission line fluxes and kinematics would help to determine whether interactions are significantly impacting the dynamical state of this system. The remaining two galaxies (COSMOS-21452 and COSMOS-13174) do not show any evidence for multi-component structure, suggesting they may host neutral gas inflows. Interestingly, both galaxies are at the peak of their star-formation histories, suggesting that the high SFRs may be fuelled by ongoing cold gas accretion.

\section{Outflow Properties}\label{sec:outflow_properties}
We have reported the first evidence for widespread \NaD\ absorption in massive (log($M_*/M_\odot >$~10) galaxies at cosmic noon, revealing that these galaxies have large neutral gas reservoirs. Approximately half of the detected absorption profiles are blueshifted, providing unambiguous evidence of neutral gas outflows. Other galaxies may have weaker outflows which are undetected at R~$\sim$~1000 due to the presence of strong ISM absorption. In this section, we investigate the properties and principal driving mechanisms of the detected neutral outflows. 

The first clue regarding the driving mechanism comes from the outflow demographics: the Blue Jay neutral gas outflows are spread almost uniformly over more than four orders of magnitude in sSFR (see Figure \ref{fig:main_sequence}). This is in tension with expectations for star-formation driven outflows, for which the incidence should increase with SFR. However, the incidence of AGN-driven ionized gas outflows at cosmic noon is observed to be independent of sSFR \citep[e.g.][]{Leung19, NMFS19}, suggesting that the neutral gas outflows we detect may be AGN-driven. 

We further explore the link between neutral outflows and AGN activity by examining galaxy emission line ratios (Section \ref{subsec:line_ratios}) and the outflow velocities, mass outflow rates and energetics (Sections \ref{subsec:vout}, \ref{subsec:mdot_calc} and \ref{subsec:energetics}, respectively). 

\subsection{Emission line ratios}\label{subsec:line_ratios}
\begin{figure}
\centering
\includegraphics[scale=0.58, clip = True, trim = 0 0 20 0]{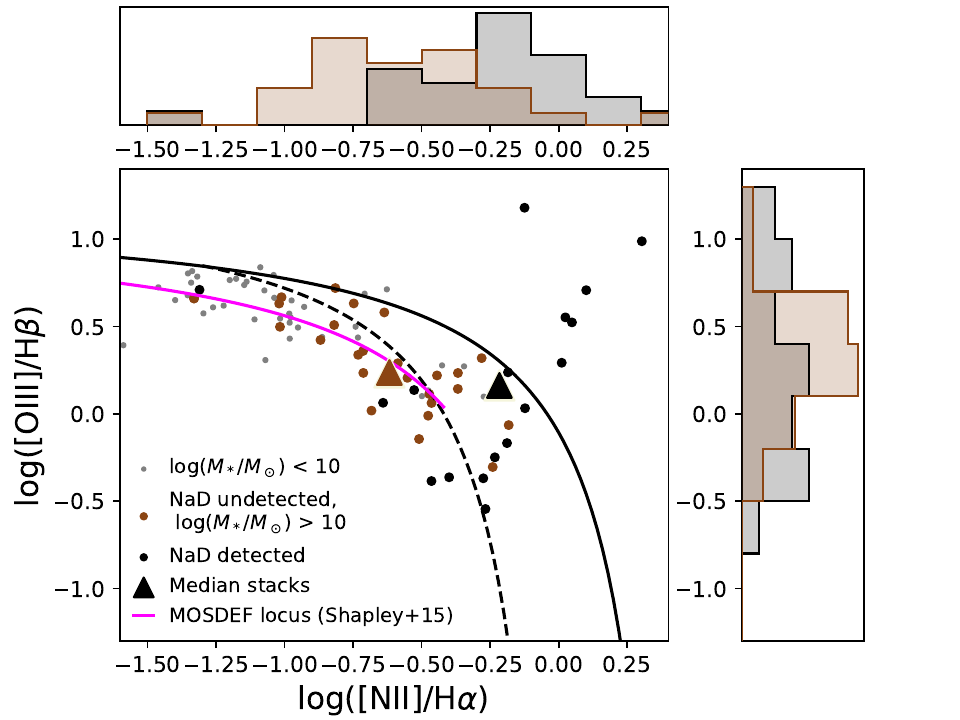}
\caption{Comparison between the emission line ratios of massive galaxies ($\log(M_*/M_\odot) >$ 10) with significant \NaD\ absorption (black), massive galaxies without \NaD\ absorption (brown), and low mass galaxies (grey). Small circles represent line ratios measured for individual galaxies with all four diagnostic emission lines detected, and the histograms show the line ratio distributions for each class of galaxies. Large triangles show line ratios measured from median stacked profiles of all galaxies in each class. The solid black curve is the theoretical upper bound for pure star-formation \citep{Ke01a} and the dashed black curve is the empirical upper bound of the $z\sim$~0 star-forming galaxy locus \citep{Ka03}. The pink curve shows the best-fit locus of $z\sim$~2.3 star-forming galaxies from the MOSDEF survey \citep{Shapley15}.}\label{fig:bpt} 
\end{figure}

Optical emission line ratios are valuable diagnostics of the principal power sources within galaxies \citep[e.g.][]{Baldry08, Veilleux87, Ke01a, Ka03}. Figure \ref{fig:bpt} shows that massive galaxies with strong interstellar \NaD\ absorption (black) have distinctly different emission line ratios from those without strong \NaD\ absorption (brown). The line ratios of individual galaxies are shown in small markers and the histograms show the line ratio distributions for the two populations. Large triangles show the ratios measured from median stacked profiles. We have verified that similar values are obtained from mean stacked profiles and by averaging the individual line ratio measurements. Galaxies lacking significant \NaD\ absorption typically lie in the star-forming and composite regions of the \NII/\Ha\ vs. \OIII/\Hb\ diagnostic diagram \citep{Baldwin81, Ke01a, Ka03} and fall close to the locus of $z\sim$~2.3 star-forming galaxies from the MOSDEF survey \citep{Shapley15}. In contrast, galaxies with detected \NaD\ absorption have significantly larger \NII/\Ha\ ratios consistent with AGN host galaxies at similar redshifts \citep[e.g.][]{Coil15}. In the local Universe, elevated \NII/\Ha\ ratios can alternatively trace shock-excitation in star-formation driven outflows \citep[e.g.][]{Sharp10}. However, star-formation driven outflows at $z\sim$~2 typically do not show very elevated \NII/\Ha\ ratios \citep[e.g.][]{Newman12_406690, Davies19, Freeman19}, perhaps because the detected line emission primarily originates from regions close to the galaxy disk where ionizing radiation from young stars dominates. Therefore, we hypothesize that strong \NaD\ absorption is preferentially associated with AGN activity. 

\subsection{Outflow Velocity}\label{subsec:vout}
Observations of outflows in ionized, molecular and neutral gas have shown that AGN-driven outflows typically have more extreme velocities than star-formation-driven outflows, where velocities $\gtrsim$~1000~\kms\ are primarily associated with AGN activity \citep[e.g.][]{Sturm11, Rupke13, Arribas14, Cicone14, Harrison16, NMFS19}. We estimate the outflow velocities for the Blue Jay targets using the velocity offset $v$ and dispersion $\sigma$ of the absorption profiles: \mbox{$v_{\rm out} = | v | + 2\sigma$}. The measured outflow velocities range from \mbox{200~--~1100~\kms}, with a median value of $\sim$~500~\kms. The fastest of the Blue Jay outflows are more likely to be AGN-driven than star-formation-driven, but the velocity information is insufficient to determine the driving mechanisms of the more moderate velocity outflows.

\subsection{Mass Outflow Rates}\label{subsec:mdot_calc}
\subsubsection{Calculations}
We estimate the neutral gas outflow rates using the time-averaged shell model presented in \citet{Rupke05a} and updated in \citet{Baron22}:
\begin{multline}\label{eqn:mdot}
\dot{M}_{\rm out} ({\rm M}_\odot {\rm yr}^{-1}) = 11.45 \, \left( C_\Omega \, \frac{C_f}{0.4} \right) \left(\frac{\rm N(HI)}{\rm 10^{21} cm^{-2}} \right) \, \times \\ \left(\frac{r_{\rm out}}{\rm 1 \, kpc} \right) \, \left( \frac{v_{\rm out}}{\rm 200 \, km \, s^{-1}} \right)
\end{multline}
where $C_\Omega$ is the large-scale covering factor related to the opening angle of the wind, \NHI\ is the hydrogen column density, $r_{\rm out}$ is the outflow radius and $v_{\rm out}$ is the outflow velocity. 

The small-scale covering fraction $C_f$ is obtained directly from the line fitting (see Equation \ref{eqn:partial_covering}). The measured values range from 0.1~--~0.9 (median 0.3); similar to what has been found in the local Universe \citep[e.g.][]{Avery22}. We assume that the outflows cover 50\% of the solid sphere (i.e. \mbox{$C_{\Omega}$ = 0.5}), consistent with the incidence of neutral outflows in local infrared galaxies \citep[e.g.][]{Rupke05b}. The geometry of neutral outflows at $z\sim$~2 is very uncertain, but the fact that we detect outflows in $\gtrsim$~25\% of massive galaxies suggests that the covering fraction cannot be much smaller than 0.25. We therefore consider the systematic uncertainty on $C_{\Omega}$ to be a factor of 2 \mbox{(i.e. $C_{\Omega}$ = 0.25~--~1)}. 

We calculate N(Na~\textsc{i}) from the optical depth at the centre of the red \NaD\ line, $\tau_{0,r}$, using the relationship from \citet{Draine11}:
\begin{multline}
\textrm{N(Na~\textsc{i})} = 10^{13} \, {\rm cm}^{-2} \left(\frac{\tau_{0,r}}{0.7580}\right) \left(\frac{0.4164}{f_{\rm lu}} \right) \times \\ \left( \frac{1215\AA}{\lambda_{\rm lu}} \right) \left( \frac{b}{10 \rm{km s}^{-1}} \right)
\end{multline}
where $f_{\rm lu}$ = 0.32 and $\lambda_{\rm lu}$ = 5897\AA\ are the oscillator strength and rest-frame wavelength of the transition, respectively, and $b$ is the Doppler parameter, equivalent to $\sqrt{2}\sigma$. By directly converting $\tau_{0,r}$ to N(Na~\textsc{i}), we are assuming that the observed absorption comes primarily from outflowing gas, with no significant contribution from gas in the ISM. The low spectral resolution of our observations means that we are unable to constrain multi-component fits allowing for contributions from both the ISM and outflows. \citet{Belli23} found that ISM gas could account for up to 44\% of the Ca~\textsc{ii k} (and by extension \NaD) absorption in COSMOS-11142. It is unlikely that the ISM component contributes more than half of the observed absorption for sources classified as outflows, because the outflow component must dominate to produce the observed negative velocity shift.

We convert N(Na~\textsc{i}) to \NHI\ assuming Milky-Way-like Na abundance and dust depletion factors, and a 10\% neutral fraction (see \citealt{Rupke05a}). This neutral fraction is based on values measured towards Milky Way stars \citep{Stokes78} and a cold extragalactic H~\textsc{i} cloud \citep{Stocke91}, and is likely to underestimate the ionization fraction in more extreme outflow environments. \citet{Baron20} measured a 5\% ionization fraction in a local AGN-driven outflow, which would increase the mass outflow rates by a factor of 2 compared to our calculations. 

The radial extent of the outflowing neutral gas cannot be measured from our observations. We estimate the likely radial extent using size measurements of 1) neutral outflows in the local Universe and 2) ionized outflows at cosmic noon. Resolved studies of \NaD\ outflows in the local Universe suggest that they typically extend a few kiloparsecs, with measured sizes ranging from \mbox{$\sim$~1~--~15~kpc} \citep[e.g.][]{Martin06, Rupke15, Rupke17, Baron20, RobertsBorsani20a, Avery22}. Similarly, \NaD\ absorption in quasar spectra is only observed within 15 kpc of galaxies \citep{Rubin22}. The two \NaD\ outflows to have been spatially analyzed at cosmic noon have sizes $\leq$~1 kpc \citep{Cresci23, Veilleux23} and 2.7 kpc \citep{DEugenio23}. In comparison, ionized gas outflows at cosmic noon typically extend to at least the galaxy effective radius, on the order of a few kpc \citep[e.g.][]{Newman12_global, Davies20, Belli23}. We conservatively adopt a 1 kpc extent and note that the mass outflow rates could be up to 10 times higher if the outflows are significantly larger than this.

When calculating the mass outflow rates we use the full Monte Carlo posterior probability distributions for $v$, $\sigma$, $C_f$ and $\tau_{0,r}$. As mentioned in Section \ref{subsec:combined_fitting}, $C_f$ and $\tau_{0,r}$ are degenerate because the \NaD\ doublet lines are blended in our observations. However, the mass outflow rate scales with the product of these two parameters (Equation \ref{eqn:mdot}), and the posterior probability distributions for the mass outflow rate are well constrained (see Appendix \ref{sec:appendix} for more details).

\renewcommand{\arraystretch}{1.25} 
\setlength{\tabcolsep}{4pt} 
\begin{table*}
\begin{tabular}{ccccccccccc}
\hline 1) Sample & 2) 3D-HST & 3) RA & 4) Dec & 5) W(NaD) & 6) $\log(M_{\rm neutral})$ & 7) $\Delta v$ & 8) $\sigma$ & 9) $v_{\rm flow}$ & 10) $\log(\dot{M}_{\rm flow})$ & 11) $\log(\eta)$ \\
ID & COSMOS ID & & & [\AA] & [M$_\odot$] & [km s$^{-1}$] & [km s$^{-1}$] & [km s$^{-1}]$ & [M$_\odot$ yr$^{-1}$] & \\ \hline
\multicolumn{11}{c}{Outflowing} \\ \hline
1 & 10565 & 10:00:22.61 & 02:17:14.16 & 4.5$^{+0.8}_{-0.8}$ & 7.85$^{+0.20}_{-0.19}$ & -148$^{+88}_{-88}$ & 225$^{+113}_{-92}$ & 602$^{+234}_{-200}$ & 1.66$^{+0.22}_{-0.31}$ & 2.55$\pm$1.19\\
2 & 11494 & 10:00:17.73 & 02:17:52.72 & 2.2$^{+0.6}_{-0.5}$ & 7.44$^{+0.23}_{-0.23}$ & -157$^{+143}_{-197}$ & 436$^{+207}_{-257}$ & 1095$^{+456}_{-657}$ & 1.51$^{+0.30}_{-0.64}$ & 1.62$\pm$1.03\\
3 & 8013 & 10:00:21.32 & 02:15:41.77 & 1.5$^{+0.9}_{-0.6}$ & 7.01$^{+0.43}_{-0.61}$ & -679$^{+133}_{-103}$ & 64$^{+99}_{-46}$ & 839$^{+203}_{-188}$ & 0.94$^{+0.51}_{-0.71}$ & 0.41$\pm$0.82\\
4 & 9871 & 10:00:18.67 & 02:16:52.09 & 11.4$^{+0.5}_{-0.4}$ & 8.01$^{+0.09}_{-0.21}$ & -174$^{+20}_{-21}$ & 91$^{+36}_{-43}$ & 354$^{+75}_{-84}$ & 1.57$^{+0.17}_{-0.33}$ & -0.01$\pm$0.47\\
5 & 10314 & 10:00:20.57 & 02:17:06.45 & 3.9$^{+0.9}_{-1.2}$ & 7.49$^{+0.17}_{-0.36}$ & -110$^{+70}_{-85}$ & 154$^{+91}_{-92}$ & 414$^{+206}_{-168}$ & 1.11$^{+0.27}_{-0.47}$ & -1.31$\pm$0.38\\
6 & 8002 & 10:00:28.78 & 02:15:39.68 & 5.8$^{+1.5}_{-1.1}$ & 8.04$^{+0.21}_{-0.23}$ & -479$^{+89}_{-114}$ & 209$^{+87}_{-94}$ & 903$^{+257}_{-265}$ & 2.01$^{+0.27}_{-0.34}$ & -0.14$\pm$0.31\\
7 & 18252 & 10:00:23.64 & 02:21:55.29 & 6.2$^{+2.6}_{-2.2}$ & 8.03$^{+0.27}_{-0.36}$ & -275$^{+247}_{-186}$ & 230$^{+193}_{-122}$ & 771$^{+382}_{-276}$ & 1.93$^{+0.34}_{-0.56}$ & -1.09$\pm$0.51\\
8 & 18668 & 10:00:31.03 & 02:22:10.43 & 9.4$^{+0.7}_{-0.7}$ & 8.06$^{+0.07}_{-0.06}$ & -143$^{+31}_{-33}$ & 228$^{+44}_{-50}$ & 600$^{+98}_{-112}$ & 1.86$^{+0.10}_{-0.13}$ & 1.64$\pm$0.87\\
9 & 11142 & 10:00:17.59 & 02:17:35.84 & 6.9$^{+0.6}_{-0.5}$ & 7.63$^{+0.19}_{-0.32}$ & -212$^{+29}_{-28}$ & 54$^{+43}_{-30}$ & 323$^{+85}_{-64}$ & 1.15$^{+0.28}_{-0.41}$ & 1.19$\pm$1.24\\
10 & 18688 & 10:00:32.04 & 02:22:13.47 & 2.0$^{+1.0}_{-0.9}$ & 7.20$^{+0.36}_{-0.54}$ & -194$^{+133}_{-130}$ & 170$^{+216}_{-119}$ & 540$^{+441}_{-239}$ & 0.93$^{+0.55}_{-0.76}$ & 0.60$\pm$1.05\\
11 & 16874 & 10:00:23.59 & 02:21:05.66 & 3.3$^{+1.0}_{-0.9}$ & 7.15$^{+0.53}_{-0.45}$ & -143$^{+74}_{-76}$ & 25$^{+88}_{-14}$ & 214$^{+171}_{-83}$ & 0.44$^{+0.80}_{-0.58}$ & -1.06$\pm$0.72\\
12 & 11136 & 10:00:26.17 & 02:17:39.58 & 10.9$^{+0.6}_{-0.6}$ & 7.68$^{+0.23}_{-0.33}$ & -118$^{+31}_{-31}$ & 43$^{+38}_{-24}$ & 209$^{+76}_{-56}$ & 1.00$^{+0.37}_{-0.45}$ & -0.88$\pm$0.46\\
13 & 10339 & 10:00:22.53 & 02:17:05.00 & 4.4$^{+0.6}_{-0.6}$ & 7.48$^{+0.22}_{-0.35}$ & -132$^{+51}_{-50}$ & 76$^{+65}_{-46}$ & 287$^{+133}_{-96}$ & 0.96$^{+0.36}_{-0.52}$ & 1.59$\pm$1.11\\
14 & 10021 & 10:00:21.45 & 02:16:56.26 & 4.0$^{+1.6}_{-1.6}$ & 7.62$^{+0.32}_{-0.50}$ & -218$^{+141}_{-120}$ & 126$^{+152}_{-86}$ & 476$^{+335}_{-195}$ & 1.29$^{+0.48}_{-0.70}$ & 0.10$\pm$0.61\\
\hline \multicolumn{11}{c}{Systemic} \\ \hline
15 & 9395 & 10:00:30.16 & 02:16:30.90 & 1.5$^{+0.4}_{-0.3}$ & 7.29$^{+0.25}_{-0.28}$ & 44$^{+137}_{-114}$ & 224$^{+185}_{-131}$ &  -- &  -- &  -- \\
16 & 7549 & 10:00:28.83 & 02:15:20.06 & 5.4$^{+1.9}_{-1.6}$ & 7.89$^{+0.23}_{-0.28}$ & -38$^{+183}_{-172}$ & 290$^{+278}_{-170}$ &  -- &  -- &  -- \\
17 & 18071 & 10:00:32.46 & 02:21:49.00 & 2.3$^{+1.0}_{-1.0}$ & 7.10$^{+0.49}_{-0.55}$ & -29$^{+99}_{-89}$ & 36$^{+108}_{-23}$ &  -- &  -- &  -- \\
18 & 19705 & 10:00:25.86 & 02:22:46.14 & 7.4$^{+3.3}_{-1.9}$ & 8.02$^{+0.32}_{-0.33}$ & -17$^{+139}_{-223}$ & 241$^{+213}_{-158}$ &  -- &  -- &  -- \\
19 & 10592 & 10:00:18.97 & 02:17:17.67 & 3.9$^{+0.7}_{-0.7}$ & 7.71$^{+0.18}_{-0.14}$ & 4$^{+92}_{-90}$ & 310$^{+131}_{-137}$ &  -- &  -- &  -- \\
20 & 16419 & 10:00:22.95 & 02:21:00.25 & 3.4$^{+1.1}_{-0.7}$ & 7.88$^{+0.19}_{-0.23}$ & -104$^{+130}_{-157}$ & 339$^{+98}_{-83}$ &  -- &  -- &  -- \\
21 & 10128 & 10:00:22.20 & 02:17:01.57 & 2.9$^{+0.4}_{-0.4}$ & 7.14$^{+0.21}_{-0.37}$ & 18$^{+45}_{-47}$ & 71$^{+58}_{-42}$ &  -- &  -- &  -- \\
22 & 17669 & 10:00:22.78 & 02:21:33.36 & 7.0$^{+2.9}_{-2.4}$ & 7.91$^{+0.38}_{-0.55}$ & 110$^{+156}_{-209}$ & 135$^{+345}_{-105}$ &  -- &  -- &  -- \\
23 & 9180 & 10:00:32.57 & 02:16:21.77 & 5.0$^{+0.8}_{-0.8}$ & 5.46$^{+1.27}_{-1.25}$ & -19$^{+41}_{-47}$ & 1$^{+13}_{-1}$ &  -- &  -- &  -- \\
24 & 10245 & 10:00:21.56 & 02:17:05.27 & 4.0$^{+0.5}_{-0.7}$ & 6.93$^{+0.37}_{-0.45}$ & 20$^{+27}_{-30}$ & 27$^{+41}_{-15}$ &  -- &  -- &  -- \\
25 & 10400 & 10:00:20.41 & 02:17:07.49 & 10.5$^{+2.1}_{-2.0}$ & 8.32$^{+0.22}_{-0.21}$ & -38$^{+244}_{-263}$ & 691$^{+172}_{-159}$ &  -- &  -- &  -- \\
\hline \multicolumn{11}{c}{BLR AGN} \\ \hline
26 & 18977 & 10:00:35.16 & 02:22:20.31 & 1.5$^{+1.0}_{-0.8}$ & 6.29$^{+1.19}_{-1.65}$ & -21$^{+262}_{-493}$ & 17$^{+254}_{-17}$ &  -- &  -- &  -- \\
27 & 12020 & 10:00:17.89 & 02:18:07.20 & 1.9$^{+0.3}_{-0.3}$ & 7.21$^{+0.18}_{-0.34}$ & -33$^{+22}_{-28}$ & 114$^{+63}_{-69}$ &  -- &  -- &  -- \\
\hline \multicolumn{11}{c}{Infalling} \\ \hline
28 & 21452 & 10:00:22.28 & 02:23:54.06 & 4.4$^{+1.5}_{-1.4}$ & 7.82$^{+0.27}_{-0.42}$ & 255$^{+145}_{-155}$ & 164$^{+121}_{-93}$ & 619$^{+218}_{-215}$ & 0.93$^{+0.61}_{-0.64}$ & -1.11$\pm$0.63\\
29 & 19572 & 10:00:31.97 & 02:22:43.40 & 5.8$^{+0.4}_{-0.4}$ & 5.59$^{+1.40}_{-1.48}$ & 27$^{+26}_{-26}$ & 1$^{+20}_{-1}$ & 40$^{+45}_{-25}$ & -2.13$^{+1.46}_{-1.34}$ & -2.50$\pm$1.57\\
30 & 13174 & 10:00:26.93 & 02:18:50.23 & 8.0$^{+1.8}_{-1.6}$ & 7.97$^{+0.32}_{-0.53}$ & 404$^{+146}_{-84}$ & 112$^{+166}_{-87}$ & 617$^{+482}_{-204}$ & 1.11$^{+0.38}_{-0.41}$ & -1.10$\pm$0.49\\
 \hline
\end{tabular}
\renewcommand{\arraystretch}{1} 
\caption{Properties of the detected \NaD\ absorption. 1) Sample ID, matching the labels in Figure \ref{fig:NaD_detections}. 2) 3D-HST COSMOS ID. 3) RA. 4) Dec. 5) Rest-frame equivalent width (\AA). \mbox{6)} Neutral gas mass, computed using Equation 3 from \citet{Baron22}. 7) Centroid velocity. 8) Velocity dispersion, assuming the nominal spectral resolution for uniform slit illumination from JDox. 9) Flow velocity, computed as $|{\Delta v}| + 2\sigma$. 10) Mass outflow rate, computed using Equation \ref{eqn:mdot}. 11) Mass loading factor, defined mass outflow rate / SFR.}\label{table:absorption_properties}
\end{table*}

\subsubsection{Results}
The measured properties of the detected \NaD\ absorption profiles and the derived neutral gas masses and outflow rates are listed in Table \ref{table:absorption_properties}. We measure outflow masses ranging from \mbox{$\log(M_{\rm out}/M_\odot)$ = 7.0 -- 8.1} (median 7.6) and mass outflow rates spanning \mbox{$\dot{M}_{\rm out}$ = 3 -- 100 M$_\odot$ yr$^{-1}$} (median \mbox{17~M$_\odot$ yr$^{-1}$}). These are consistent with neutral outflow properties measured for star-forming and AGN host galaxies in the local Universe \citep[e.g.][]{Rupke05b, Cazzoli16, Baron20, RobertsBorsani20, Avery22} as well as at $z\sim$~2 \citep{Perna15, Cresci23}. The neutral gas outflow rates are also comparable to the ionized gas outflow rates measured for galaxies at similar stellar mass and redshift \citep[e.g.][]{NMFS19}. We emphasize that the estimates presented here are based on conservative assumptions for the outflow extent and Na ionization fraction, and the true outflow rates could plausibly be an order of magnitude larger. In the two \mbox{$z\sim$~2~--~3} galaxies for which neutral and ionized mass outflow rates have been directly compared, the neutral outflow rates exceed the ionized outflow rates by approximately a factor of 100 \citep{Belli23, DEugenio23}. Our results emphasize that it is important to account for the neutral phase in order to paint a complete picture of ejective feedback. 

\begin{figure*}
\centering
\includegraphics[scale=1, clip = True, trim = 10 160 0 0]{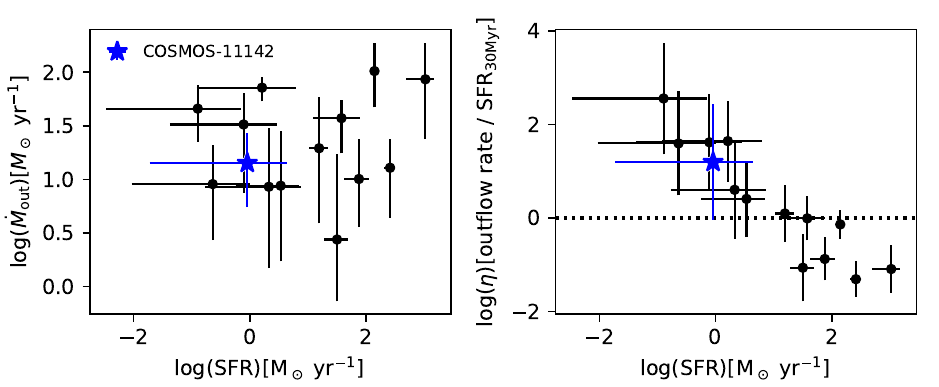}
\caption{Mass outflow rate ($\dot{M}_{\rm out}$, left) and mass loading factor ($\eta$, right) as a function of SFR. There is no correlation between $\dot{M}_{\rm out}$ and SFR, and as a consequence, $\eta$ differs strongly between the star-forming and quenching populations. In high SFR galaxies (log(SFR)[$M_\odot$ yr$^{-1}$] $\gtrsim$ 0), $\eta \lesssim$~1, consistent with expectations for self-regulating star formation feedback. However, the quenching galaxies have mass loading factors of \mbox{4~--~360}, suggesting that AGN activity is required to power the outflows in these systems.} \label{fig:mass_loading} 
\end{figure*}

Interestingly, we do not find any correlation between neutral gas outflow rate and galaxy SFR, as shown in the left-hand panel of Figure \ref{fig:mass_loading}. As a consequence, the outflow mass loading factors ($\eta$, defined as mass outflow rate divided by SFR) differ strongly between the star-forming and quenching populations (Figure \ref{fig:mass_loading}, right). The outflows launched from the most actively star-forming galaxies \mbox{($\log$(SFR)[M$_\odot$ yr$^{-1}$] $\gtrsim$ 0)} have mass-loading factors of $\eta \lesssim$~1, consistent with expectations for star-formation-driven outflows \citep[e.g.][]{Finlator08, Dave11, Somerville15}. In contrast, the mass loading factors for the lower SFR galaxies range from 4~--~360. It is unlikely that energy injection by young stars could remove gas so much faster than the stars themselves are forming. However, many of the low SFR galaxies in our sample show strong Balmer absorption lines indicative of a recent rapid decline in SFR. We therefore investigate whether it is possible that the outflows were launched during a recent starburst phase, in which case the high mass loading factors could be an artefact of the time delay between the launch of the outflows and the SFR measurement (which is averaged over the last 30 Myr). The slowest outflow in our sample has a velocity of 210~\kms, and absorption line measurements have found that \NaD\ absorption is only observed within 15~kpc of galaxies \citep[e.g.][]{Rubin22}. This corresponds to a maximum reasonable outflow travel time of 70 Myr. We compute mass-loading factors using SFRs in different age bins from the \textsc{Prospector} fitting. Mass-loading factors of order unity are only found when using SFRs more than 100 Myr in the past. This is at least 50\% longer than the maximum reasonable outflow travel time, suggesting that past star-formation could not reasonably have powered these outflows and providing further evidence that they are driven by AGN activity.

\subsection{Energy and momentum rates}\label{subsec:energetics}
Next, we investigate whether the current levels of star-formation and AGN activity in the Blue Jay galaxies are sufficient to explain the energetics of the neutral gas outflows. Figure \ref{fig:outflow_energetics} compares the kinetic energy and momentum rates of the outflows ($\dot{E}_{\rm out}$ and $\dot{p}_{\rm out}$, respectively) with the rates of energy and momentum injection by supernovae and AGN. For the supernovae, we adopt the mechanical energy and momentum rate scalings from \citet{Veilleux05} based on solar metallicity Starburst99 models \citep{Leitherer99}: \mbox{$\dot{E}_{\rm SN}$ = 7~$\times$~10$^{41} \times$~SFR[$M_\odot$ yr$^{-1}$] erg/s}, \mbox{$\dot{p}_{\rm SN}$ = 5~$\times$~10$^{33} \times$~SFR[$M_\odot$ yr$^{-1}$] dyne}. The AGN bolometric luminosity is estimated from the \OIII\ luminosity applying a bolometric correction factor of 600 \citep{Netzer09}. The \OIII\ luminosity is corrected for extinction using the median $A_V$ from the \textsc{Prospector} posterior probability distribution (including extra attenuation towards towards young stars), and the uncertainty on $A_V$ is propagated through to the uncertainty on the \OIII\ luminosity. The \OIII\ emission in most of the outflow host galaxies is dominated by AGN activity (see Figure \ref{fig:bpt}), and \citet{Netzer09} show that even for composite galaxies where more than half of the Balmer line emission is due to star-formation, the total \OIII\ luminosity predicts the bolometric luminosity to within a factor of 2. Energy conserving AGN-driven outflows are expected to have kinetic energy rates equivalent to 5\% of the AGN bolometric luminosity \citep[see][and references therein]{King15}. The AGN momentum flux output is \mbox{$\dot{p}_{\rm AGN}$ = $L_{\rm AGN}/c$}, but in energy-conserving outflows, the momentum outflow rate can be boosted by a factor of $\sim$~5-20 due to entrainment of ISM gas in the wind \citep[e.g.][]{Faucher12}. To account for this, we plot lines for outflow momentum rates equivalent to $L_{\rm AGN}/c$ and $20 \times L_{\rm AGN}/c$.

The top row of Figure \ref{fig:outflow_energetics} compares the outflow energetics with predictions for star-formation-driven outflows. In the most actively star-forming galaxies \mbox{($\log(L_{\rm SF})$[erg s$^{-1}$] $\gtrsim$ 44)}, the energy injected by star-formation is likely sufficient to power the observed outflows. However, in the lower SFR systems, the energy (momentum) injection rates are up to 60 (280) times larger than the predicted inputs. It is unlikely that this discrepancy can be explained by systematic uncertainties on the mass outflow rates. The shaded grey regions show the range of plausible values accounting for uncertainties on the ISM absorption contribution, ionization fraction and wind opening angle (see Section \ref{subsec:mdot_calc}). We adopt a very conservative outflow size of \mbox{$r_{\rm out}$ = 1 kpc}, and assuming a larger extent would only increase the outflow energy further above the energy and momentum injection by supernovae. This, together with the implausibly large mass-loading factors (see Section \ref{subsec:mdot_calc}), provides strong evidence to suggest that the neutral gas outflows from the low SFR galaxies are unlikely to be powered by star-formation. 

The bottom row of Figure \ref{fig:outflow_energetics} compares the outflow energetics with predictions for AGN-driven outflows. We see that in all cases, the AGN are powerful enough to drive the observed outflows.

\section{Discussion}\label{sec:discussion}
\begin{figure*}
\centering
\includegraphics[scale=1, clip = True, trim = 0 10 0 0]{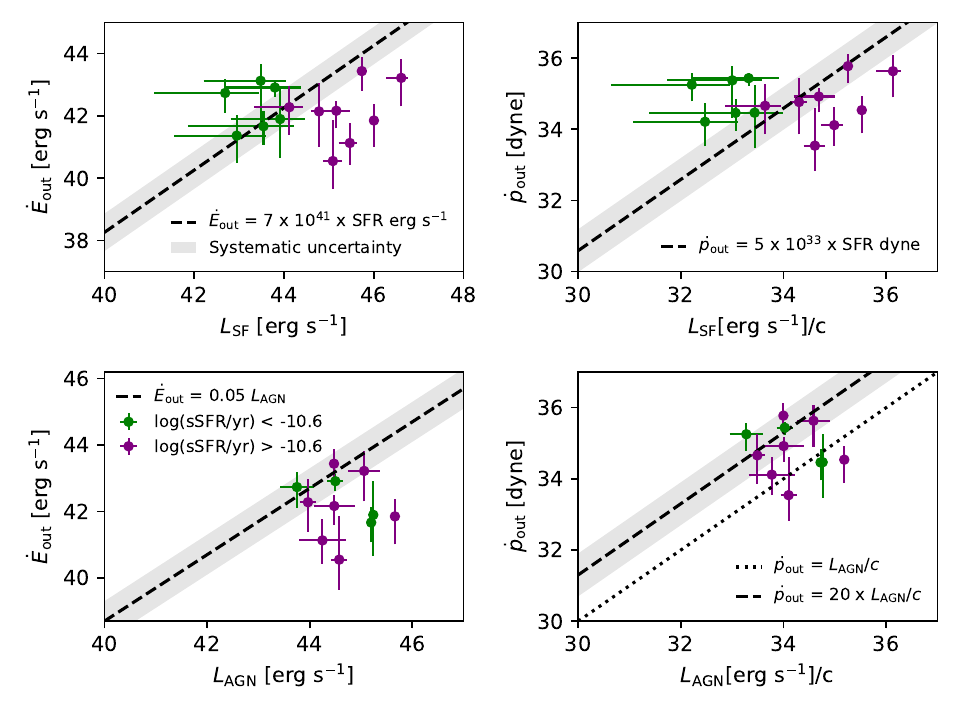}
\caption{Measured energy (left) and momentum (right) rates of the outflows compared to the luminosity and momentum flux from young stars (top) and AGN (bottom). Dashed lines indicate the expected energy and momentum injection into the outflows. Shaded bands represent a factor of four uncertainty in either direction, accounting for potential ISM \NaD\ absorption and/or incorrect assumptions about the wind opening angle or ionization fraction (see discussion in Section \ref{subsec:mdot_calc}). The shaded region does not account for variations in outflow extent which could increase the outflow energy and momentum by up to an order of magnitude (moving the data points up). The outflows from the low SFR galaxies (log(L$_{\rm SF}$)[erg s$^{-1}$] $\lesssim$ 44) are too powerful to be driven by star-formation, but could easily be powered by the observed AGN activity.} \label{fig:outflow_energetics} 
\end{figure*}

Our investigation of the outflow driving mechanisms indicates that AGN activity likely plays a major role in powering the observed neutral gas outflows in massive $z\sim$~2 galaxies. The incidence of neutral outflows is independent of (s)SFR (Figure \ref{fig:main_sequence}). Galaxies with strong \NaD\ absorption show high \NII/\Ha\ ratios consistent with AGN ionization, whereas galaxies without \NaD\ absorption show lower \NII/\Ha\ ratios consistent with photoionization by young stars (Figure \ref{fig:bpt}). Some outflows have velocities exceeding 500~\kms\ (Section \ref{subsec:vout}). The case for AGN-driven outflows is particularly strong for low SFR galaxies where the mass loading factors range from \mbox{4~--~360} (Figure \ref{fig:mass_loading}) and the outflows are removing energy and momentum tens to hundreds of times faster than they can be injected by young stars (Section \ref{fig:outflow_energetics}).

In summary, the neutral gas outflows in the Blue Jay sample are evenly distributed across star-forming and quenching galaxies, and AGN accretion appears to play a major role in driving these outflows. This is in contrast to the local Universe where the majority of neutral outflows are found in star-forming galaxies and are consistent with being star-formation-driven \citep[e.g.][]{Rupke05c, Chen10, Bae18, Concas19, Nedelchev19, RobertsBorsani19, Avery22}. There is some evidence that outflows in local low SFR galaxies are preferentially associated with AGN activity \citep[e.g.][]{Concas19, RobertsBorsani19, Avery22, Sun23}, consistent with our findings. The role of AGN in driving outflows may be enhanced in our sample because we are probing significantly brighter AGN: the median AGN luminosity of the Blue Jay outflow hosts (6~$\times$~10$^{44}$~erg/s) is about two orders of magnitude higher than that of optically selected samples at $z\sim$~0 \citep[e.g.][]{RobertsBorsani19, Avery22}; consistent with the known redshift evolution in AGN luminosity \citep[e.g.][]{Rosario12, Carraro20}. In the local Universe, neutral outflows from quasar host galaxies are faster and have higher mass outflow rates than outflows from star-forming galaxies \citep[e.g.][]{Veilleux05, Rupke13, Cazzoli16}, suggesting that luminous AGN play a significant role in driving outflows at all redshifts.

Neutral outflows from low sSFR galaxies are much more prevalent at cosmic noon than in the local Universe. This may be because low sSFR galaxies at $z\sim$~2 have had a lot less time to grow and quench than their $z\sim$~0 counterparts, and as a result they have much younger stellar populations \citep[e.g.][]{Belli19, Carnall19, Tacchella22}. 85\% of the massive, low sSFR galaxies in the Blue Jay sample have light-weighted ages less than 1 Gyr (Park et al., in prep) and could therefore be post-starburst galaxies. The molecular gas reservoirs of post-starburst galaxies have been observed to decline with time since quenching \citep[e.g.][]{French18, Bezanson22, Baron23}, making it less likely to observe neutral gas outflows from the most evolved sources (although some older post-starburst galaxies do have detectable quantities of molecular gas; e.g. \citealt{Rowlands15}). Both the incidence and velocity of \NaD\ outflows in local massive galaxies appear to decrease as galaxies age and quench \citep[e.g.][]{Sun23}. We do not find any evidence for a correlation between outflow velocity and either SFR or light-weighted age in our sample; however, it is likely that even if such a correlation was present, it would be masked due to the small sample size and relatively large measurement errors.

The rapid depletion of the molecular gas reservoirs is thought to be driven by powerful outflows, which have been observed in many post-starburst galaxies (both at $z\sim$~0 and $z\sim$~1; e.g. \citealt{Tremonti07, Davis12, Alatalo15, Maltby19, Baron20, Baron22}). The low sSFR outflow host galaxies in our $z\sim$~2 sample may similarly trace a `blowout' phase where strong AGN-driven outflows are ejecting large amounts of cold gas, leading to rapid quenching of star-formation. This picture is supported by detailed analyses of two post-starburst galaxies at $z\sim$~2~--~3 \citep{Belli23, DEugenio23}. Both galaxies experienced a burst of star-formation \mbox{0.3~--~0.8 Gyr} ago followed by a rapid decline in star-formation activity. These galaxies host powerful, AGN-driven neutral gas outflows that are ejecting cold gas 10~--~100 times faster than it can be converted into stars. The rapid quenching of these systems may therefore be fully explained by ejective AGN feedback. 

We have shown that similarly powerful neutral gas outflows are prevalent across the massive galaxy population at cosmic noon. The outflows from the quenching galaxies in our sample have mass-loading factors of \mbox{4 -- 360} (see Figure \ref{fig:mass_loading}), consistent with the case studies above. Our work indicates that AGN-driven neutral gas outflows may represent a dominant avenue for fast quenching at $z\sim$~2. Chemical evolution modelling of massive quiescent galaxies at $z\sim$~1 has shown that mass-loading factors of order 10 are required to explain the stellar magnesium abundances, suggesting that these galaxies experienced powerful outflows  prior to quenching \citep[e.g.][]{Leethochawalit19, Zhuang23}. Many high redshift quiescent galaxies show emission line ratios consistent with AGN ionization (e.g. \citealt{Belli17b, Newman18, Belli19, Park22}, Bugiani et al., in prep), providing additional evidence that AGN feedback plays a crucial role in quenching star-formation. 

It is important to note that only a small fraction of the outflowing gas we detect may be able to escape the galaxy halos. The halo escape velocity is expected to be 2.5~--~3 times the galaxy circular velocity $v_{\rm circ}$ \citep[e.g.][]{Weiner09, Swinbank19}. We are not able to measure circular velocities directly because we only have slit spectra, so we adopt $v_{\rm circ} \sim$~300~\kms\ which is typical of AGN host galaxies at this redshift \citep{NMFS19}. This corresponds to halo escape velocities of \mbox{$v_{\rm esc, halo}$~=~750~--~900~\kms}. Only 4/14 (29\%) of the outflows in our sample exceed these velocities, suggesting that the majority of gas ejected in the outflows will remain in the halo and may eventually be re-accreted onto the galaxies. It is possible that the slowest outflows in our sample may not even escape from their host galaxies. However, it is difficult to robustly determine the fate of the outflowing gas for several reasons: 1) the errors on the outflow velocity measurements in Table \ref{table:absorption_properties} are quite large (the median error is $\sim$~200~\kms), 2) the measured velocity dispersions are likely under-estimated given the compact nature of our targets (see footnote \ref{footnote:spectral_res}), meaning that the outflow velocities are also under-estimated, 3) the calculated outflow velocity depends on the assumed geometry and definition for $v_{\rm out}$, and 4) the escape velocity depends on the galactocentric radius which is unknown. Furthermore, \citet{Sun23} found that outflows diminish with age, meaning that low SFR galaxies may have driven much faster outflows in the past.

Star-formation quenching likely involves a combination of gas removal and heating. \citet{Roy21} found that a large fraction of radio-detected quiescent galaxies show infalling neutral gas probed by redshifted \NaD\ absorption, and the authors suggest that radio jets may be responsible for heating the accreted gas and preventing it from forming new stars. Slow outflows may similarly inhibit star-formation through turbulence or redistribution of gas within the galaxies \citep[see also][]{Luo22, Sun23}. The powerful ejection of gas through outflows is crucial to explain the observed rapid quenching of galaxies in the early Universe \citep[e.g.][]{Belli19, Park22}, whilst maintenance mode feedback is required to prevent rejuvenation and keep galaxies quiescent over long timescales.

\section{Summary and conclusions}\label{sec:conclusion}
We have used JWST/NIRSpec observations of 113 galaxies at \mbox{1.7 $< z <$ 3.5} selected from the mass-complete Blue Jay survey to investigate the demographics and properties of neutral gas outflows, traced by \NaD\ absorption, at cosmic noon. Our observations have revealed for the first time that interstellar \NaD\ absorption is widespread in massive ($\log(M_*/M_\odot) >$~10) galaxies at $z\sim$~2. Our main findings are as follows:
\begin{itemize}
\item We detect interstellar \NaD\ absorption in 30/113 galaxies. The detections are almost exclusively associated with massive ($\log(M_*/M_\odot) >$ 10) galaxies, for which the detection fraction is 46\%. Lower mass galaxies likely have insufficient columns of gas and dust to shield \NaD\ against ionization.
\item 50\% of the \NaD\ absorption profiles are blueshifted by at least 100~\kms, providing unambiguous evidence for neutral gas outflows. These neutral outflows are observed across the entire massive galaxy population, with similar incidence rates in star-forming and quenching galaxies. 
\item 39\% of the \NaD\ profiles are consistent with the galaxy systemic velocity. These primarily trace cool gas in the ISM, but may also have weaker underlying outflow components that are hidden at R~$\sim$~1000.
\item 3 galaxies (11\%) show redshifted absorption profiles indicative of infalling gas. Of these, one galaxy shows a complex morphology and emission-line ratios consistent with shock excitation, suggesting that the redshifted absorption may trace bulk flows of gas within an interacting system. The other two galaxies appear isolated and are at the peaks of their star-formation histories, suggesting that their star-formation may be fuelled by ongoing accretion of cool gas.
\item Assuming a conservative outflow extent of 1~kpc, we compute neutral mass outflow rates of \mbox{3 -- 100~$M_\odot$~yr$^{-1}$}. These are comparable to or greater than ionized gas outflow rates previously reported for other galaxies with similar stellar masses and redshifts. Existing measurements of neutral gas outflow sizes range from \mbox{1~--~15 kpc}, so the true outflow extents and mass outflow rates from the Blue Jay galaxies could plausibly be up to an order of magnitude larger than we report.
\item Multiple lines of evidence indicate that the outflows are likely to be AGN-driven. Galaxies with strong interstellar \NaD\ absorption have enhanced \NII/\Ha\ ratios indicative of AGN activity. The outflow incidence does not depend on the level of star-formation activity. Star-formation cannot power the outflows from the low SFR galaxies, where the outflow mass loading factors range from \mbox{4~--~360} and the energy and momentum outflow rates exceed the injection rates from supernovae by at least an order of magnitude. 
\item The presence of strong neutral outflows in quenching systems could indicate that they are undergoing a post-starburst `blowout' phase powered by the AGN. The outflow velocities range from \mbox{200~--~1100~\kms}, albeit with large uncertainties (typical error 200~\kms). It is difficult to robustly constrain the fate of the outflowing gas, but in most cases the estimated outflow velocities are lower than the expected halo escape velocities, suggesting that the bulk of the outflowing material will remain in the galaxy halos. Nevertheless, the strong AGN-driven ejection of cold gas provides a mechanism to explain the rapid quenching of star-formation in massive quiescent galaxies in the early Universe. Maintenance mode feedback (e.g. through radio jets) may also be required to prevent the re-accretion of cold gas and keep the galaxies quiescent.
\end{itemize}

Our results indicate that powerful, AGN-driven neutral gas outflows are prevalent across the massive galaxy population at $z\sim$~2 and are likely to be a dominant channel for fast quenching at this epoch. 

\section*{Acknowledgements}
We thank the anonymous referee for their suggestions which improved the clarity of this manuscript. We thank Karl Glazebrook for thought-provoking discussions. RLD is supported by the Australian Research Council Centre of Excellence for All Sky Astrophysics in 3 Dimensions (ASTRO 3D), through project number CE170100013. RLD is supported by the Australian Research Council through the Discovery Early Career Researcher Award (DECRA) Fellowship DE240100136 funded by the Australian Government. SB is supported by the the ERC Starting Grant ``Red Cardinal'', GA 101076080. RE acknowledges the support from grant numbers 21-atp21-0077, NSF AST-1816420, and HST-GO-16173.001-A as well as the Institute for Theory and Computation at the Center for Astrophysics. RW acknowledges funding of a Leibniz Junior Research Group (project number J131/2022).

This work is based on observations made with the NASA/ESA/CSA James Webb Space Telescope. The data were obtained from the Mikulski Archive for Space Telescopes at the Space Telescope Science Institute, which is operated by the Association of Universities for Research in Astronomy, Inc., under NASA contract NAS 5-03127 for JWST. These observations are associated with program GO 1810. The Blue Jay Survey is funded in part by STScI Grant JWST-GO-01810. This work also makes use of observations taken by the 3D-HST Treasury Program (GO 12177 and 12328) with the NASA/ESA HST, which is operated by the Association of Universities for Research in Astronomy, Inc., under NASA contract NAS5-26555. 

\section*{Data Availability}

The JWST/NIRSpec MSA spectra used in this paper were obtained through the Cycle 1 program Blue Jay (GO 1810, PI Belli) and are publicly available for download from MAST.


\bibliographystyle{mnras}
\bibliography{mybib}



\appendix

\section{Outflow parameter constraints}\label{sec:appendix}
\begin{figure*}
\centering
\includegraphics[scale=0.5]{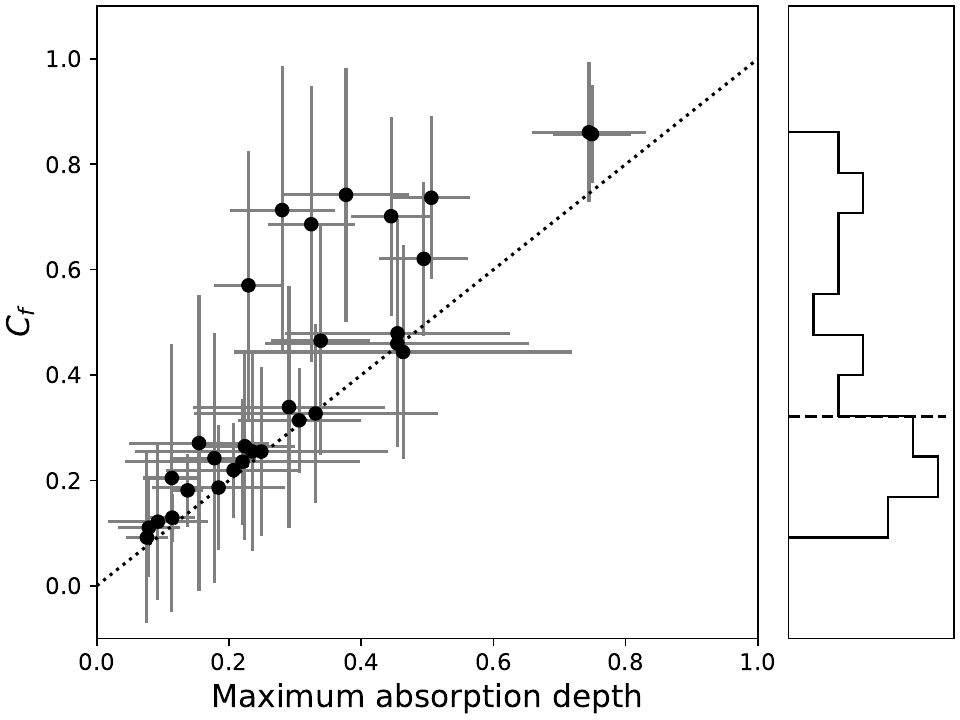} \hspace{5pt} \includegraphics[scale=0.3, clip = True, trim = 0 5 0 25]{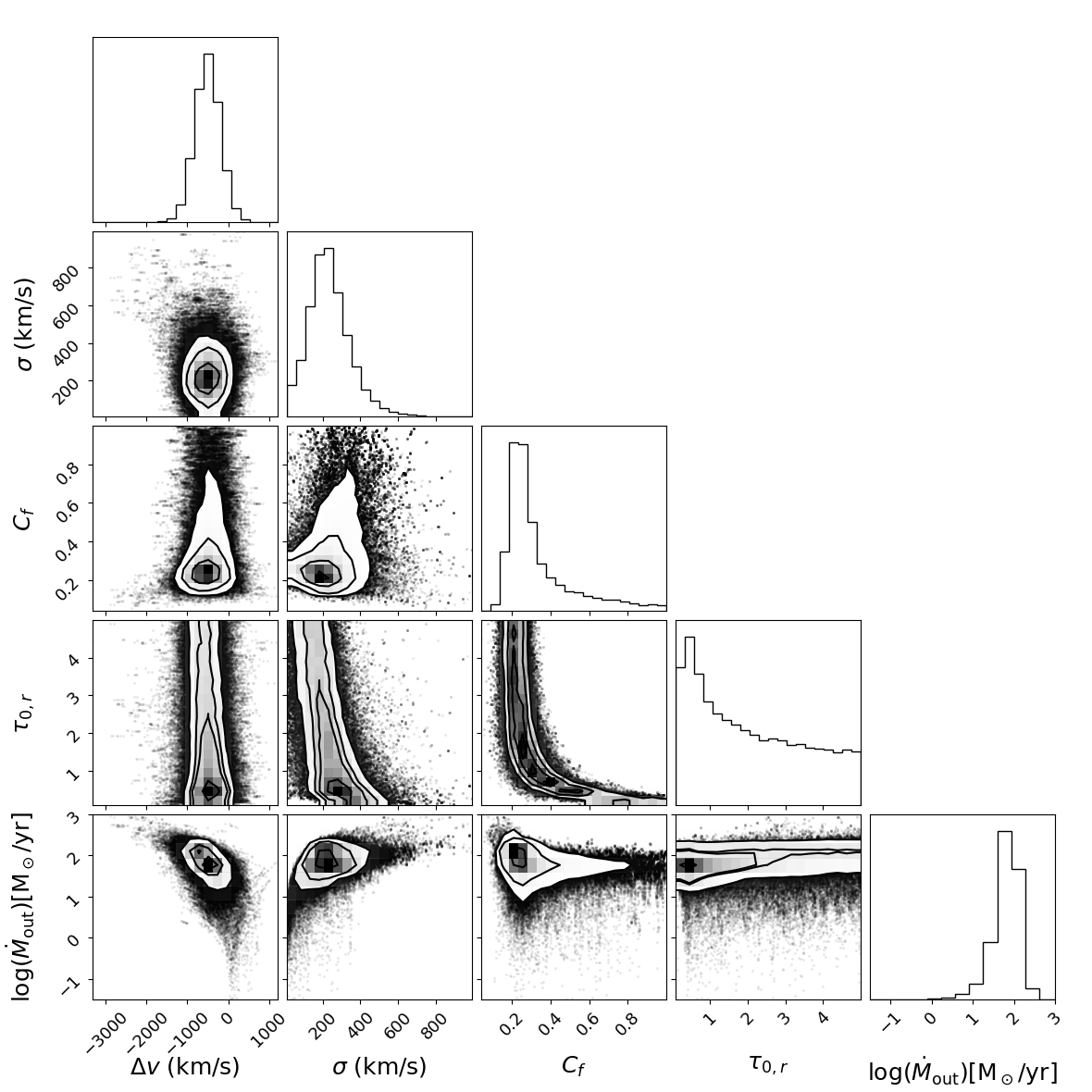}
\caption{Left: Outflow covering fraction $C_f$ as a function of the measured maximum absorption depth. The dotted line indicates a 1:1 relation. The maximum absorption depth provides a lower limit on $C_f$. Right: Single and joint posterior probability distributions for the \NaD\ absorption parameters (velocity offset $\Delta v$, dispersion $\sigma$, covering fraction $C_f$, and optical depth $\tau_{0,r}$) as well as the mass outflow rate $\dot{M}_{\rm out}$ for COSMOS-10565. $C_f$ and $\tau_{0,r}$ are degenerate but $\dot{M}_{\rm out}$, which scales with the product of these two parameters, is well constrained.} \label{fig:corner} 
\end{figure*}

We fit the \NaD\ absorption profiles using a partial covering model parametrized by the gas covering fraction $C_f$, optical depth $\tau$, velocity $v$ and dispersion $\sigma$ (Equation \ref{eqn:partial_covering}). The optical depth modulates the shape and depth of the absorption profile as well as the relative strength of the red and blue doublet lines. The covering fraction also impacts the depth of the observed absorption. At low spectral resolution, these parameters become degenerate \citep[e.g.][]{Rupke05b}, raising the question of how well the total mass outflow rate can be constrained. 

From Equation \ref{eqn:partial_covering}, we see that the maximum absorption depth is at most 1 - $C_f$. In other words, if the absorption depth is 80\% (with absorption reaching down to 20\% of the continuum level), it implies \mbox{$C_f \geq$ 0.8}. The left-hand panel of Figure \ref{fig:corner} shows $C_f$ as a function of the maximum absorption depth, with the dotted line indicating a 1:1 relation. All points lie on or above the 1:1 line, as expected from Equation \ref{eqn:partial_covering}. 

The right-hand panel of Figure \ref{fig:corner} shows single and joint posterior probability distributions for the four outflow model parameters and the derived mass outflow rate for COSMOS-10565. Focusing on $C_f$, we see that the lower boundary is well constrained by the maximum absorption depth, with a long tail towards larger values. The optical depth $\tau_{0,r}$ is very poorly constrained but varies inversely with $C_f$ because both parameters impact the absorption depth. The mass outflow rate scales with the product of $C_f$ and $\tau_{0,r}$ (Equation \ref{eqn:mdot}), and because these parameters are inversely dependent, the mass outflow rate is well constrained. 


\bsp	
\label{lastpage}
\end{document}